\documentclass[onecolumn]{revtex4-2}
\usepackage{amsmath}
\usepackage{amssymb}
\usepackage{amsthm}
\usepackage{bm}
\usepackage[shortlabels]{enumitem}
\usepackage{tabularx}
\usepackage{graphicx}
\usepackage{tikz-cd}
\usepackage{xcolor}
\usepackage{hyperref}
\usepackage[normalem]{ulem}

\usepackage[english]{babel}
\usepackage[letterpaper,top=2cm,bottom=2cm,left=3cm,right=3cm,marginparwidth=1.75cm]{geometry}

\begin{document}

\title{Local precursors to anomalous dissipation in Navier-Stokes turbulence: Burgers vortex-type models and simulation analysis}
\author{Georgy Zinchenko, Vladyslav Pushenko and J\"org Schumacher}
\affiliation{Institute of Thermodynamics and Fluid Mechanics, Technische Universität Ilmenau, P.O.Box 100565, D-98684 Ilmenau, Germany}
\date{\today}

\begin{abstract}
Anomalous dissipation is a dissipation mechanism of kinetic energy which is established by a sufficiently spatially rough velocity field. It implies that the rescaled mean kinetic energy dissipation rate becomes constant with respect to Reynolds number ${\rm Re}$, the dimensionless parameter that characterizes the strength of turbulence, given that ${\rm Re}\gg 1$. The present study aims at bridging this statistical behavior of high-Reynolds-number turbulence to specific structural building blocks of fluid turbulence -- local vortex stretching configurations which take in the simplest case the form of Burgers' classical vortex stretching model from 1948. We discuss the anomalous dissipation in the framework of Duchon and Robert for this analytical solution of the Navier-Stokes equations, apply the same analysis subsequently to a generalized model of randomly oriented Burgers vortices by Kambe and Hatakeyama, and analyse finally direct numerical simulation data of three-dimensional homogeneous, isotropic box turbulence in this respect. We identify local high-vorticity events in fully developed Navier-Stokes turbulence that approximate the analytical models of strong vortex stretching well. They also correspond to precursors of enhanced anomalous dissipation.   
\end{abstract}
\maketitle

\section{Introduction}
In a fully developed three-dimensional turbulent Navier-Stokes flow ${\bm u}$, kinetic energy is dissipated at small scales due to the molecular kinematic viscosity $\nu$ of the fluid. This dissipation implies a conversion of turbulent kinetic energy density into heat. It is considered as a consequence of a continued transfer of turbulent kinetic energy from the larger scales of the flow across an inertial subrange to the smaller ones where the velocity field is spatially smooth \cite{Kolmogorov1941,Frisch1995,Davidson2004,Dubrulle2019}.
The average rate at which this dissipation occurs is the mean kinetic energy dissipation rate $\langle\epsilon\rangle$ which equals the mean transfer rate of turbulent kinetic energy across the inertial range in the statistically steady regime. It is given as an ensemble average of the turbulent kinetic energy dissipation rate field $\epsilon({\bm x},t)$. For a homogeneous turbulent velocity field ${\bm u}({\bm x},t)$ follows
\begin{equation}
\langle \epsilon \rangle = \nu \langle |{\bm \nabla}{\bm u}|^2\rangle\,. 
\label{eq:diss1}
\end{equation}
The {\em dissipative anomaly} states that the rescaled mean kinetic energy dissipation stays constant as the kinematic viscosity of the fluid tends towards zero
\begin{equation}
\lim_{\nu\to 0}\frac{\langle \epsilon \rangle L_0}{U^3_{\rm rms}} = \mbox{const}\quad\mbox{with}\quad U_{\rm rms}=\sqrt{\langle {\bm u}^2\rangle}\,. 
\label{eq:diss2}
\end{equation}
Here, $L_0$ is an outer scale of turbulence, e.g., the integral length scale \cite{Davidson2004} and $U_{\rm rms}$ is the root mean square velocity, a measure of the strength of velocity fluctuations. This behavior is counter-intuitive at first glance because one might expect that as viscosity decreases towards zero (or the Reynolds number ${\rm Re}=U_{\rm rms}L_0/\nu$ increases towards infinity) the mean kinetic energy dissipation rate would also decrease. However, the strong dissipative anomaly implies that in the limit of zero viscosity, the flow is still dissipating energy at a non-zero rate caused by a sufficiently spatially rough velocity field in the inertial subrange. Spatial roughness is characterized by a H\"older exponent $h$ which determines a power law dependence of velocity increments with respect to a spatial distance, $|{\bm u}({\bm x}+{\bm \xi})-{\bm u}({\bm x})|\sim \xi^h$ with $\xi=|{\bm \xi}|$. Sufficient roughness implies a Hölder exponent  of $h<1/3$; in contrast a spatially smooth velocity field results to $h=1$. 

This idea dates back to 1949 when Lars Onsager suggested that sufficiently rough {\em weak solutions} of the three-dimensional ideal Euler equations may be important for Navier-Stokes turbulence at very high Reynolds numbers. Onsager  formulated his anomalous dissipation hypothesis for an incompressible Euler flow at $\nu=0$ \cite{Onsager1949}, see also \cite{Eyink2024,Sreenivasan2024}. It was proven for exponents $h<1/3$ in a series of works by De Lellis and Sz\'{e}kelyhidi starting with \cite{Delellis2009} for $h<1/10$ and finally by Isett \cite{Isett2018} for $h<1/3$. The dissipative anomaly is also a fundamental aspect of Navier-Stokes turbulence in the large-Reynolds number limit, sometimes denoted as the zeroth law of turbulence. Empirical evidence for its existence has been collected from several experiments and direct numerical simulations (DNS) starting with Sreenivasan \cite{Sreenivasan1984} and updated in subsequent works \cite{Sreenivasan1998,Kaneda2003} It was also shown to exist in the bulk of turbulent convection flows \cite{Pandey2022}. 

In this work, we want to take a different approach to the dissipative anomaly. The goal is to connect this fundamental statistical property to existing flow structures in fluid turbulence. One essential structural building block of the kinetic energy cascade in three-dimensional fluid turbulence is the local stretching of vortices, a mechanism which was first described by Jan Burgers in an analytical solution to the Navier-Stokes equations \cite{Burgers1948} and subsequently analysed in turbulent flows \cite{Townsend1951,Ashurst1987,Ohkitani1994,Hamlington2008a,Hamlington2008}. We therefore apply the spatial filtering over a scale $\varepsilon$ which was originally suggested by Duchon and Robert to study the dissipative anomaly \cite{Duchon2000}. The application of the variable-scale filtering (or the joint integration with a $C^{\infty}$ test function in a distributive sense) for each term of the Navier-Stokes equations results in an additional anomalous dissipation term in the local kinetic energy balance. See also ref. \cite{Dubrulle2019} for a comprehensive discussion. Here, we calculate this term analytically by applying Gaussian wavelets as filter kernels or test functions \cite{Mallat1999}. Furthermore, we repeat this analysis for an extension of the original Burgers case to an ensemble of randomly oriented Burgers vortices which was suggested by Hatakeyama and Kambe \cite{Hatakeyama1997,Kambe2000}. Finally, we use our own direct numerical simulation (DNS) data of homogeneous isotropic box turbulence to identify {\em local} high vorticity events and reveal their role as {\em local precursors} to anomalous dissipation. The latter step is motivated to previous experimental studies in von K\'{a}rm\'{a}n swirling flows \cite{Dubrulle2019,Saw2016}. We show that in all three cases {\em precursors} to anomalous dissipation can be identified. We wish to stress at the end of this paragraph that precursors to anomalous dissipation can be considered only since finite Reynolds numbers and thus a finite Kolmogorov dissipation length $\eta_K=(\nu^3/\langle\epsilon\rangle)^{1/4}$ are here obtained. In other words, the joint limit of vanishing kinematic viscosity and vanishing filter scale cannot be executed in the following. 

The outline of the manuscript is as follows. In Sec. II, we will present the framework of Duchon and Robert in brief and provide the Burgers vortex solution for completeness. Section III discusses the anomalous dissipation in the Duchon-Robert picture for the steady Burgers vortex analytically. We extend this model and determine the anomalous dissipation term for the random Burgers vortex case of Kambe and Hatakeyama in Sec. IV. Finally, we use our own DNS data of three-dimensional homogeneous, isotropic turbulence to identify local intense Burgers vortex stretching events in Sec. V. We finish the work with a summary and an outlook in Sec. VI.

\section{Methods}
\subsection{Duchon-Robert framework}
In the following, we present in brief the framework of Robert and Duchon for the dissipative anomaly for an incompressible three-dimensional Navier-Stokes flow \cite{Duchon2000}. The Navier-Stokes (NS) equations for the components of the velocity field ${\bm u}=(u_x,u_y,u_z)$ and the spatial coordinates ${\bm x}=(x,y,z)$ are given by
\begin{align}
\label{NS_eq}
\partial_t {\bm u} + \partial_j (u_j{\bm u})  &= -\frac{1}{\rho} {\bm \nabla} p + \nu \Delta {\bm u}, \\
\label{C_eq}
{\bm \nabla} \cdot {\bm u} &= 0,
\end{align}
where $p$ is the pressure field of the fluid, $\rho$ is the (constant) mass density of the fluid, and $\nu$ is the kinematic viscosity of the fluid, which follows from the dynamic viscosity $\mu$ by $\nu=\mu/\rho$. The index $j=x,y,z$ and the Einstein summation convention is used. The variable $u_j$ is the $j$-th component of ${\bf u}$ and $\partial_j$ the partial derivative with respect to the $j$-th spatial coordinate.

Let $\varphi\in C^{\infty}(V)$ be any infinitely differentiable function with a compact support and $\int_V \varphi dV=1$ with flow volume $V$ (which will be a 3-torus for simplicity). Let us also define the following rescaled function for the spatial coordinates ${\bm \xi}=(\xi_x,\xi_y,\xi_z)$ which is given by
\begin{equation}
    \varphi^\varepsilon({\bm \xi})=\frac{1}{\varepsilon^3}\varphi\left(\frac{{\bm \xi}}{\varepsilon}\right)\,,
\end{equation}
where $\varepsilon$ serves as a filtering or coarse-graining scale. By denoting the convolutions ${\bm u}^\varepsilon = \varphi^\varepsilon * 
{\bm u},\ p^\varepsilon = \varphi^\varepsilon * p,\ \left(u_j{\bm u}\right)^\varepsilon = \varphi^\varepsilon * \left(u_j{\bm u}\right)$, we can regularize eq. \eqref{NS_eq} to
\begin{equation}
    \partial_t {\bm u}^\varepsilon + \partial_j ( u_j {\bm u})^\varepsilon = -\frac{1}{\rho} {\bm \nabla} p^\varepsilon + \nu \Delta {\bm u}^\varepsilon 
\label{NS_eq1}
\end{equation}
A component-wise multiplication of \eqref{NS_eq1} with ${\bm u}$, a further component-wise multiplication of \eqref{NS_eq} with ${\bm u}^\varepsilon$, and an addition of both gives the following local kinetic energy balance equation 
\begin{equation}
\label{regularized_kin_en_eq_1}
    \partial_t ({\bm u} \cdot {\bm u}^\varepsilon) + {\bm u} \cdot \partial_j (u_j {\bm u}^\varepsilon) - {\bm u} \cdot \partial_j (u_j {\bm u}^\varepsilon) + {\bm \nabla} \cdot \left[ {\bm u}({\bm u} \cdot {\bm u}^\varepsilon) + \frac{{\bm u} p^\varepsilon + {\bm u}^\varepsilon p}{\rho}\right] = \nu \Delta ({\bm u} \cdot {\bm u}^\varepsilon) - 2\nu {\bm \nabla} {\bm u} : {\bm \nabla} {\bm u}^\varepsilon.
\end{equation}
Furthermore, one gets
\begin{equation}
    \frac{1}{2} \int{{\bm \nabla} \varphi^\varepsilon({\bm \xi}) \cdot \delta {\bm u}^2 \delta {\bm u}\, d^3\xi} = 
    -\frac{1}{2} \partial_j(u_j {\bm u}^2)^\varepsilon 
    +{\bm u} \cdot \partial_j (u_j {\bm u})^\varepsilon
    + \frac{1}{2} u_j \partial_j ({\bm u}^2)^\varepsilon 
    - u_j {\bm u} \cdot \partial_j {\bm u}^\varepsilon\,.
\label{NS_eq2}    
\end{equation}
with $\delta {\bm u}({\bm \xi})={\bm u}({\bm x}+{\bm \xi})-{\bm u}({\bm x})$ the velocity increments over a distance vector ${\bm \xi}$. We will use frequently the shorter-hand notation of $\delta {\bm u}$ to ease the longer mathematical expressions. Substituting the second and fourth terms on the right hand side of \eqref{NS_eq2} into \eqref{regularized_kin_en_eq_1} and dividing the whole resulting equation by 2 leads to
\begin{align}
\label{regularized_kin_en_eq_2}
    \partial_t \frac{{\bm u} \cdot {{\bm u}}^{\varepsilon}}{2} &+ {\bm \nabla} \cdot \left[ {\bm u} \frac{{\bm u} \cdot {{\bm u}}^{\varepsilon}}{2} + \frac{{\bm u}{p}^{\varepsilon} + {{\bm u}}^{\varepsilon}p}{2\rho} + \frac{1}{4} \left( {\bm u} {\bm u}^2 \right)^\varepsilon - \frac{1}{4} {\bm u} \left( {\bm u}^2 \right)^\varepsilon \right] \\
    & = \nu \Delta \frac{{\bm u} \cdot {{\bm u}}^{\varepsilon}}{2} - \nu {\bm \nabla} {\bm u} : {\bm \nabla} {{\bm u}}^{\varepsilon} - \frac{1}{4} \int {\bm \nabla} \varphi^{\varepsilon} \left( {\bm \xi} \right) \cdot \delta {\bm u}^2 \delta {\bm u} \; d^3 \xi\,.\nonumber
\end{align}
If we take the limit $\varepsilon\to 0$ then several terms in \eqref{regularized_kin_en_eq_2} will simplify. The term $({\bm u} \cdot {{\bm u}}^{\varepsilon})/2$ converges for example to the kinetic energy $E$. In detail we get,
\begin{align}
    \underset{\varepsilon \to 0}{\mathop{\lim }}\frac{{\bm u}\cdot {\bm u}^\varepsilon }{2} =\frac{{\bm u}^2}{2}=E,\;\;
    \underset{\varepsilon \to 0}{\mathop{\lim }} \left({\bm u}p^\varepsilon+{\bm u}^\varepsilon p\right)=2{\bm u}p,\;\;
    \underset{\varepsilon \to 0}{\mathop{\lim }}\left[ \left( {\bm u} {\bm u}^2 \right)^\varepsilon -  {\bm u} \left( {\bm u}^2 \right)^\varepsilon \right]=0,\;\;
    \underset{\varepsilon \to 0}{\mathop{\lim }}{\bm \nabla} {{\bm u}}^{\varepsilon}={\bm \nabla} {\bm u}\,. \nonumber
\end{align}
For this limit $\varepsilon\to 0$, we can consequently rewrite eq. \eqref{regularized_kin_en_eq_2} to the following local kinetic energy balance
\begin{equation}
\label{kin_en_eq}
    \partial _t E+{\bm \nabla} \cdot \left( {\bm u}E+{\bm u}\frac{p}{\rho } \right)=\nu \Delta E-\nu {{\left( {\bm \nabla} {\bm u} \right)}^{2}}-\underset{\varepsilon \to 0}{\mathop{\lim }}D(\varepsilon,{\bm r})\,
\end{equation}
The additional dissipation term $D$ on the right hand side is defined as follows,
\begin{equation}
\label{diss_term}
    D(\varepsilon,{\bm x}):=\frac{1}{4}\int{{\bm \nabla} {{\varphi }^{\varepsilon }}({\bm \xi})\cdot \delta {\bm u} (\delta {\bm u})^2\,{{d}^{3}}\mathbf{\xi }}\,.
\end{equation}
This is the anomalous dissipation term which does not vanish for a sufficient roughness of the velocity field in the inertial subrange, i.e. for scales $\eta_K\ll \xi=|{\bm \xi}|\ll L_0$ and $\delta{\bm u}\cdot{\bm e}_{\xi}\sim \xi^h$ with $h<1/3$. Here, $L_0$ the integral length scale, and ${\bm e}_{\xi}={\bm \xi}/\xi$. We proceed with the Burgers vortex solution.

An additional note on eq. \eqref{kin_en_eq} is appropriate here. The kinetic energy dissipation rate field due to kinematic viscosity is given here as $\epsilon({\bm x},t)=\nu ({\bm \nabla}{\bm u})^2$ and not as $\epsilon({\bm x},t)=(\nu/2)({\nabla}{\bm u}+({\bm \nabla}{\bm u})^T)^2$. The former is sometimes denoted as pseudo-dissipation rate field \cite{Pope2000} and differs by an additional term that vanishes only when an averaging is done for homogeneous turbulence. This however does not alter the anomalous dissipation term $D$ which is in the focus of our analysis.

\subsection{Burgers vortex solution}
\begin{figure}[h]
\centering
    \includegraphics[width=0.8\linewidth]{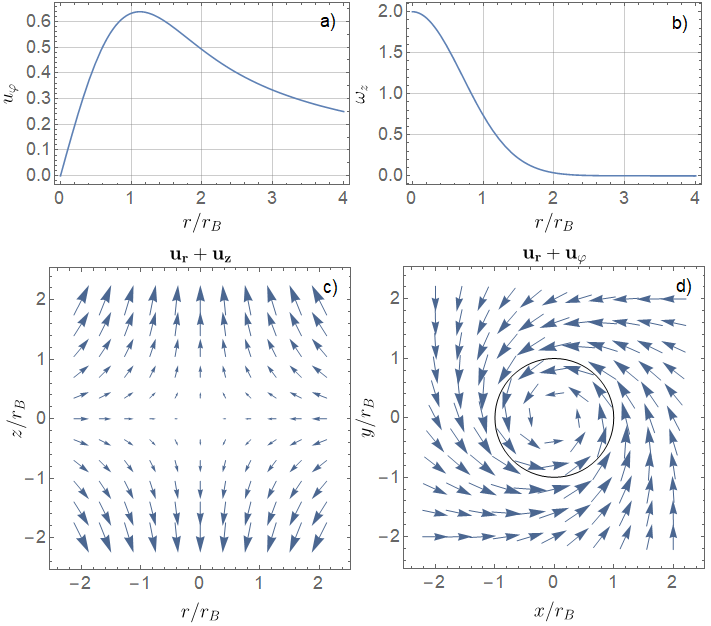}
    \caption{Burgers vortex solution. (a) Azimuthal velocity distribution in radial direction. (b) Vorticity distribution in radial direction. (c) Velocity vector field in vertical symmetry plane. (d) Velocity vector field in horizontal plane at $z=0$.}
    \label{Velocity_distrib_BV}
\end{figure}

\begin{figure}[h]
    \centering
\includegraphics[width=0.9\linewidth]{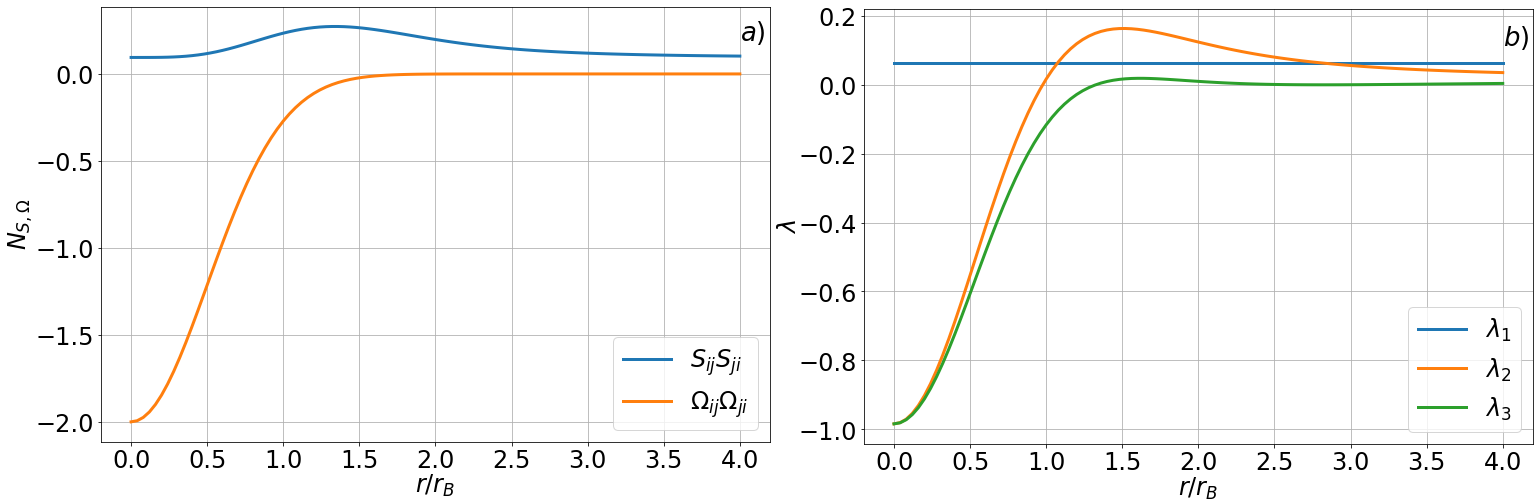}
\caption{Velocity gradients of Burgers vortex solution. (a) Norms of the symmetric rate of strain and antisymmetric vorticity tensors \eqref{norms_stress_tensor}. (b) Distribution of three eigenvalues in radial direction \eqref{lambdas}. The vortex core is defined by $\lambda_2<0$.}
\label{norms_lambdas}
\end{figure}

The Burgers vortex \cite{Burgers1948} is an exact and steady solution to the Navier–Stokes equations governing a viscous incompressible flow, see also \cite{Davidson2004} for a time-dependent extension of this model. It can be considered as a fundamental building block of turbulence which describes the vortex stretching in a three-dimensional pure strain flow in combination with the generation of a self-induced strain that is connected to component $u_{\varphi}(r)$ and opposes the original continued stretching by the background flow \cite{Townsend1951,Hamlington2008a}. Its velocity and vorticity components are given in axisymmetric form by
\begin{equation}
\label{Velocity_BV}
    {{u}_{r}}=-\alpha_s r,\quad {{u}_{z}}=2\alpha_s z, \quad {{u}_{\varphi }}=\frac{\Gamma }{2\pi r}\left( 1-\exp \left( -\frac{\alpha_s {{r}^{2}}}{2\nu } \right) \right), \quad
     {{\omega }_{z}}=\frac{\Gamma }{2\pi }\frac{\alpha_s }{\nu }\exp \left( -\frac{\alpha_s {{r}^{2}}}{2\nu } \right)\,.
\end{equation}
Here, $\alpha_s$ is the strain rate of the background flow and $\Gamma$ is the prescribed circulation of the vortex. The characteristic length is given by the Burgers radius,
\begin{equation}
  r_B=\sqrt{\frac{2\nu }{\alpha_s }},  \quad  
 u_0=\frac{\Gamma }{2\pi r_B},  \quad
 \omega_0=\frac{u_0}{r_B}. 
 \label{Dimesionless_parameters}
\end{equation}
In the limit of very large Reynolds number, which corresponds $\nu\to 0$, the Burgers radius $r_B\to 0$ which implies that an increasing vorticity $\omega_z$ is concentrated increasingly closer to the $z$-axis given that the circulation $\Gamma$ remains constant. The dimensionless form results to 
\begin{equation}
\label{Velocity_dimless_BV}
    u_r = -\beta_v r, \quad u_z = 2\beta_v z, \quad u_\varphi = \frac{1-\exp(-r^2)}{r}, \quad \omega_z = 2\exp(-r^2),
\end{equation}
where the dimensionless parameter $\beta_v =\alpha_s r_B/{u}_{0}$ is defined by the strain rate, circulation, Burgers radius and the kinematic viscosity. The parameter $\beta_v$ is an inverse Reynolds number, $\beta_v = 2/Re_B$ with $Re_B=u_0r_B/\nu$, see next subsection. Resulting velocity components amplitudes and vector fields are shown in Fig. \ref{Velocity_distrib_BV}.
In order to determine the core of the Burgers vortex, the $\lambda_2$-criterion \cite{Davidson2004} is used in the following. It consists of several steps. First, the velocity gradient tensor is obtained as
\begin{equation}
    J=\left( \begin{matrix}
   {{J}_{rr}} & {{J}_{\varphi r}} & 0  \\
   {{J}_{r\varphi }} & {{J}_{\varphi \varphi }} & 0  \\
   0 & 0 & {{J}_{zz}} 
\end{matrix} \right)=\left( \begin{matrix}
   \frac{\partial {{u}_{r}}}{\partial r} & -\frac{{{u}_{\varphi }}}{r} & 0  \\
   \frac{\partial {{u}_{\varphi }}}{\partial r} & \frac{{{u}_{r}}}{r} & 0  \\
   0 & 0 & \frac{\partial {{u}_{z}}}{\partial z} 
\end{matrix} \right)\,. \nonumber
\end{equation}
Then the symmetric rate of strain and the antisymmetric vorticity tensors are found to
\begin{equation}
    S=\frac{J+{{J}^{T}}}{2},\quad \Omega =\frac{J-{{J}^{T}}}{2}.\nonumber
\end{equation}
Note that the self-induced strain due to vortex stretching is given by the off-diagonal component $S_{r\varphi}=(J_{r\varphi}+J_{\varphi r})/2$ \cite{Hamlington2008a}. It is specifically this term which leads to a non-vanishing kinetic energy dissipation (calculated by an integral over the whole volume) in the limit $\nu\to 0$ as already pointed out by Burgers in the concluding remarks of his work \cite{Burgers1948}. The norms of both tensors are given by
\begin{equation}
\label{norms_stress_tensor}
    N_{S}={{S}_{ij}}{{S}_{ji}},\quad N_{\Omega }={{\Omega }_{ij}}{{\Omega }_{ji}}.
\end{equation}
For the $\lambda_2$-criterion, three eigenvalues of the new tensor $\Lambda=S^2+\Omega^2$ have to be determined. The new tensor is given by
\begin{equation}
    \Lambda =\frac{1}{2}\left( \begin{matrix}
   2J_{rr}^{2}+2{{J}_{\varphi r}}{{J}_{r\varphi }} & \left( {{J}_{rr}}+{{J}_{\varphi \varphi }} \right)\left( {{J}_{\varphi r}}+{{J}_{r\varphi }} \right) & 0  \\
   \left( {{J}_{rr}}+{{J}_{\varphi \varphi }} \right)\left( {{J}_{\varphi r}}+{{J}_{r\varphi }} \right) & 2J_{\varphi \varphi }^{2}+2{{J}_{\varphi r}}{{J}_{r\varphi }} & 0  \\
   0 & 0 & 2J_{zz}^{2} 
\end{matrix} \right)
\label{lambda_matrix}
\end{equation}
The matrix determinant of \eqref{lambda_matrix} is equal to zero for three following eigenvalues $\lambda_{1,2,3}$,
\begin{equation}
\label{lambdas}
  {{\lambda }_{1}}=J_{zz}^{2}, \quad
 {{\lambda }_{2,3}}=\left( J_{rr}^{2}+{{J}_{\varphi r}}{{J}_{r\varphi }} \right)\pm {{J}_{rr}}\left( {{J}_{\varphi r}}+{{J}_{r\varphi }} \right).  
\end{equation}
Their distribution in the Burgers vortex is shown in Fig. \ref{norms_lambdas}
An isocontour of the vorticity magnitude would be cylinder surfaces with a certain radius $r/r_B$ centered about the $z$-axis, see also the sketch in Fig. \ref{BV_representation}.

\section{Anomalous dissipation for aligned Burgers vortex}
\begin{figure}[h]
    \centering
    \includegraphics[width=0.5\linewidth]{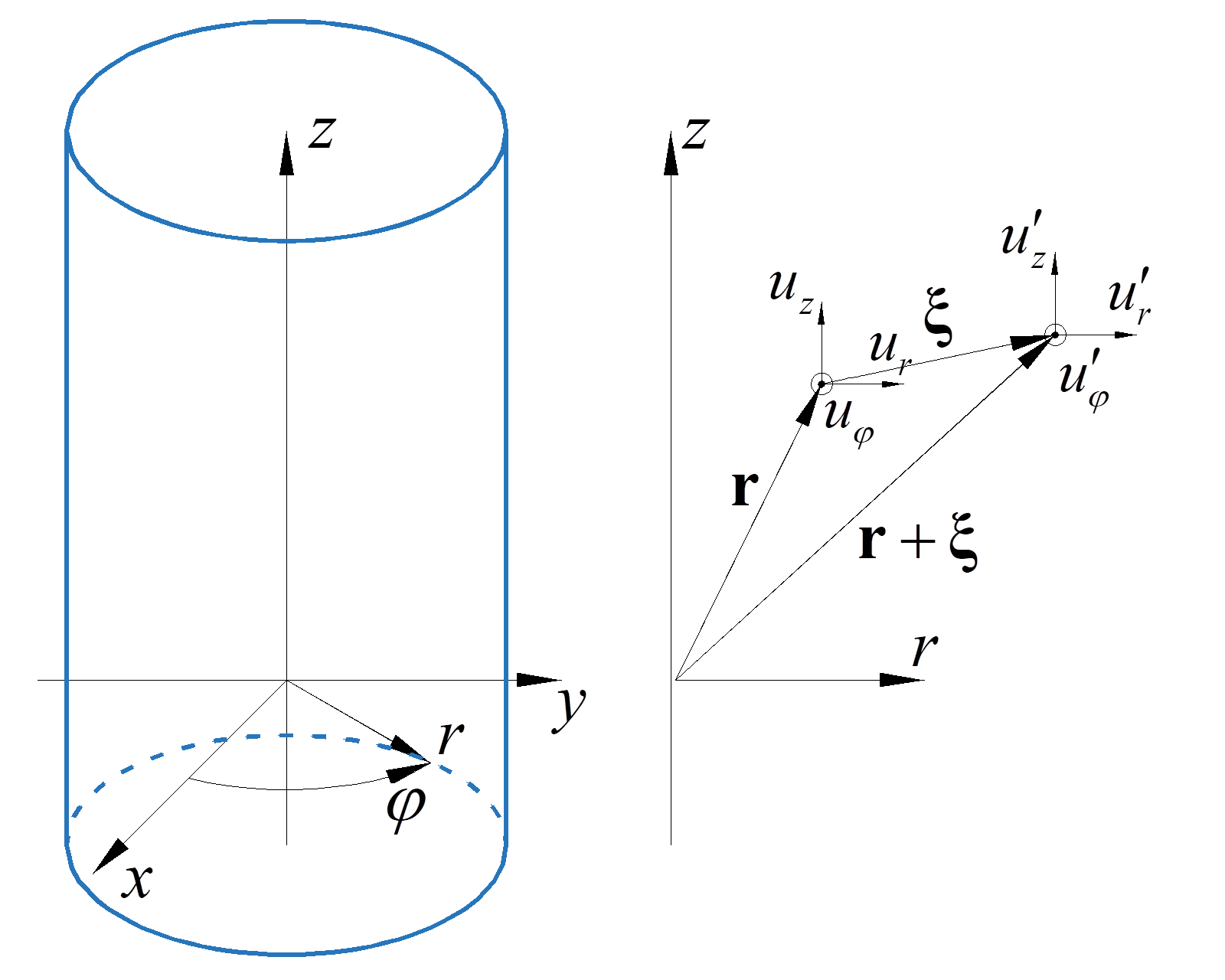}
    \caption{Left: Burgers vortex tube in a cylindrical coordinate system. Right: Detailed coordinate system. Vector ${\bm r}=\left(r,z\right)$ represents the reference point, vector ${\bm \xi}=\left(\xi_r,\xi_z\right)$ represents the increment vector. The velocity increment is here defined as $\delta {\bm u}({\bm \xi})={\bm u}\left({\bm r}+{\bm \xi}\right)-{\bm u}\left({\bm r}\right)$.}
    \label{BV_representation}
\end{figure}

\begin{figure}[h]
\centering
    \includegraphics[width=0.95\linewidth]{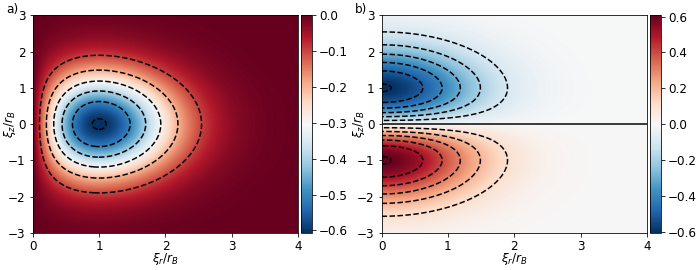}
    \caption{Spatial structure of the 1st-order vector-valued wavelet ${\bm \psi}_1$. (a) Contour plot of the radial component of the wavelet. (b) Contour plot of the vertical component of the wavelet.}
\label{wavelet_distrib}
\end{figure}

\begin{figure}[h]
\centering
\includegraphics[width=0.95\linewidth]{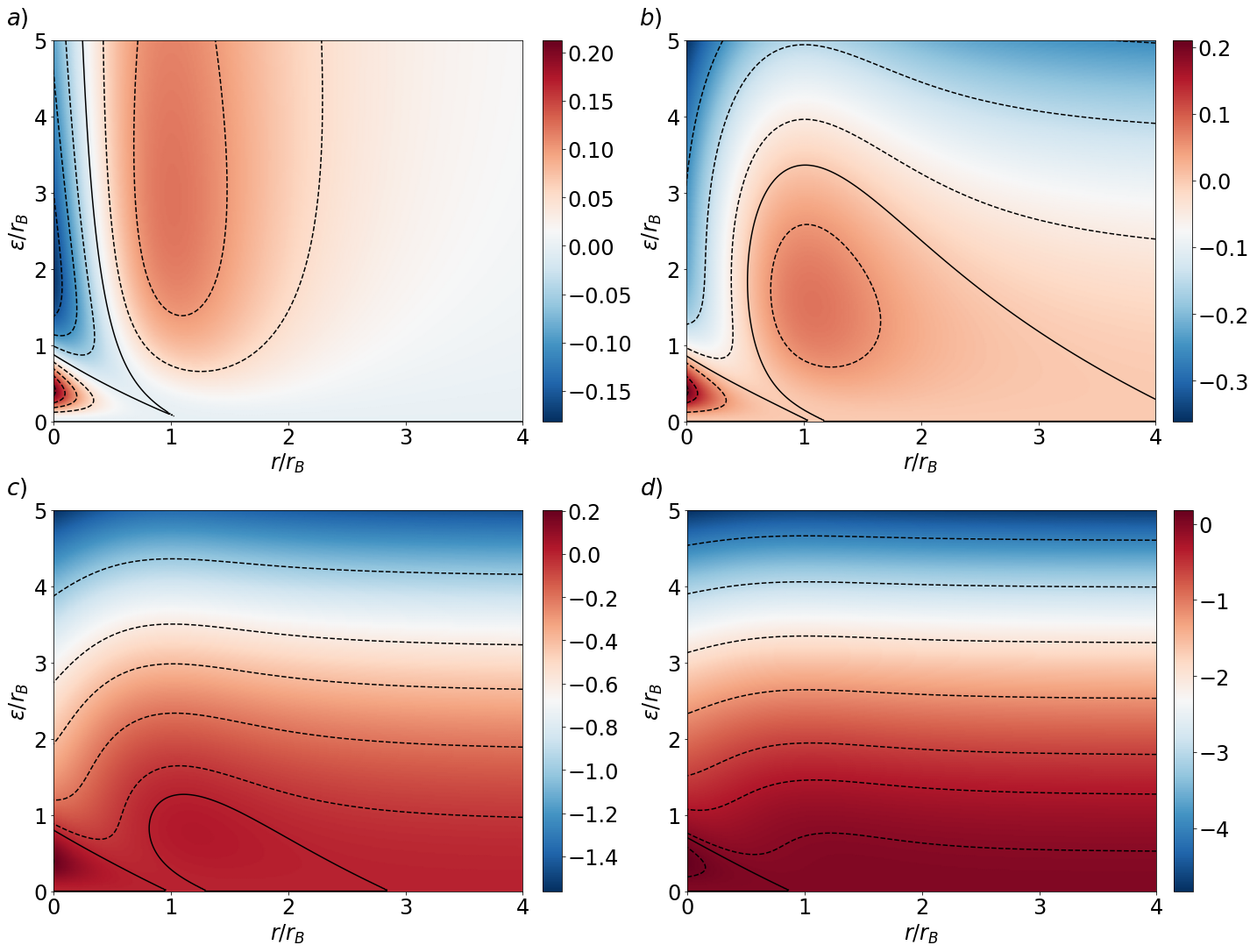}
\caption{(a) Contour plot for the rotational anomalous dissipation term $D_{\rm rot}(\varepsilon,r)$ given by \eqref{diss_term_rot}. (b) Total anomalous dissipation term $D(\varepsilon,r)=D_{\rm irr}(\varepsilon)+D_{\rm rot}(\varepsilon,r)$ in $\varepsilon$--$r$ space with parameter $\beta_v=0.03$. (c) Same as (b) for $\beta_v=0.07$. (d) Same as (b) for $\beta_v=0.125$ respectively. All scales are normalized by the Burgers radius. The solid line is the zero-contour-level in all panels.}
\label{Diss_term_contour}
\end{figure}

For the Burgers vortex configuration that is aligned with the $z$ axis and which is shown in Fig. \ref{BV_representation}, we apply the kinetic energy equation \eqref{kin_en_eq} and calculate the anomalous dissipation term \eqref{diss_term}. By using the dimensionless parameters \eqref{Dimesionless_parameters} and the characteristic pressure $p_0=\rho u_0^2$, we can rewrite the kinetic energy equation in dimensionless form to
\begin{equation}
    \beta_v {{\partial }_{t}}\frac{{{{\bm u}}^{2}}}{2}+{\bm \nabla} \cdot \left( {\bm u}\frac{{{{\bm u}}^{2}}}{2}+{\bm u}p \right)=-\underset{\varepsilon \to 0}{\mathop{\lim }}\,\frac{1}{4{{\varepsilon }^{4}}}\int{{{{\bm \psi}}_{1}}\left( \frac{{\bm \xi }-{\bm s}}{\varepsilon } \right)\cdot \delta {{{\bm u}}^{2}}\delta {\bm u}\,{{d}^{3}}\xi }+\frac{1}{\operatorname{Re}}\left( \Delta \frac{{{{\bm u}}^{2}}}{2}-{{\left( {\bm \nabla} {\bm u} \right)}^{2}} \right),
\label{Dimensionless_NS_eq}
\end{equation}
where the Burgers vortex Reynolds number is given by
\begin{equation}
{\rm Re}_B=\frac{r_Bu_0}{\nu }\,.
\end{equation}
The dimensionless test function $\psi_1$ in \eqref{Dimensionless_NS_eq} is selected using the theory of wavelets \cite{Mallat1999}. Here $\varepsilon$ is a scale factor (or coarse graining length) that works as a spatial filter for kinetic energy transfer. The coordinate $\mathbf{s}$ denotes a shift of the signal. In the present case, we use vector-valued Gaussian wavelets which are defined as follows
\begin{equation}
\label{Gaussian_wavelet}
    {{\bm \psi}_{m}}({\bm \xi})={{(-1)}^{m+1}}{{\bm \nabla }^{m}}\exp ( -{{{\bm \xi}}^{2}}/2 )\quad\mbox{for}\quad m=1,2,3...,
\end{equation}
where ${\bm \xi}^2=\xi_x^2+\xi_y^2+\xi_z^2=\xi_r^2+\xi_z^2$ in Cartesian coordinates and cylindrical coordinates. To calculate the dissipation term in \eqref{Dimensionless_NS_eq}, we have to use 1st-order Gaussian wavelet, i.e. a wavelet of the following form 
\begin{equation}
    {{\bm \psi}}_{1}({\bm \xi})={\bm \nabla} \exp(-{\bm\xi }^{2}/2)=-{{\xi }_{r}}\exp(-{{\bm \xi }}^{2}/2){\bm e}_r-{{\xi }_{z}}\exp(-{\bm \xi}^{2}/2){\bm e}_z
\label{1_order_Gaussian_wavelet}
\end{equation}
The components $\psi_r$ and $\psi_z$ of this wavelet are illustrated in Fig. \ref{wavelet_distrib}. In order to calculate the anomalous dissipation term $D(\varepsilon,\bm{r})$, we have to integrate over all possible increment vectors in cylindrical coordinates. This leads to
\begin{equation}
\label{diss_tem_calc1}
    D(\varepsilon,{\bm r}) = \frac{1}{4\varepsilon^4}\int\limits_{-\infty}^{\infty}\int\limits_{0}^{2\pi}\int\limits_{0}^{\infty} {\bm \psi}_1\left(\frac{{\bm \xi}-{\bm s}}{\varepsilon}\right)\cdot {\bm S}({\bm r},{\bm \xi})\xi_r \,d\xi_r \,d\xi_\varphi \,d\xi_z,\quad\mbox{with}\quad {\bm S}({\bm r},{\bm \xi})=(\delta{\bm u})^2\delta {\bm u} \,.
\end{equation}
Note that the increment is now $\delta{\bm u}={\bm u}({\bm r}+{\bm \xi})-{\bm u}({\bm r})$, see also Fig. \ref{BV_representation}. In \eqref{diss_tem_calc1} neither the test function nor the signal depend on the azimuthal coordinate $\xi_{\varphi}$. Thus we simplify the expression to 
\begin{equation}
\label{diss_tem_calc2}
    D(\varepsilon,{\bm r}) = \frac{\pi}{2\varepsilon^4}\int\limits_{-\infty}^{\infty}\int\limits_{0}^{\infty} {\bm \psi}_1\left(\frac{{\bm \xi}-{\bm s}}{\varepsilon}\right)\cdot {\bm S}({\bm r},{\bm \xi})\xi_r \,d\xi_r \,d\xi_z.
\end{equation}
For a better understanding of the dissipation term behaviour as a function of the scale factor $\varepsilon$, we use the following substitution for the increment variable ${\bm x}=({\bm \xi }-{\bm s})/\varepsilon$, where the vector ${\bm x}=(x_r,x_z)$ has again 2 components only. As a result \eqref{diss_tem_calc2} transforms to
\begin{equation}
\label{diss_tem_calc3}
    D(\varepsilon,{\bm r}) = \frac{\pi}{2\varepsilon^2} \int\limits_{-\infty}^{\infty}\int\limits_{-s_r/\varepsilon}^{\infty} {\bm \psi}_1({\bm x})\cdot {\bm S}({\bm r},\varepsilon{\bm x}+{\bm s})(\varepsilon x_r+s_r)\,d{x_r}\,d{x_z}\,.
\end{equation}
For our purposes, the signal shift, that is used in the theory of wavelet transforms, can be removed, i.e. we take ${\bm s}=0$. Then the dissipation term \eqref{diss_tem_calc3} simplifies further to
\begin{equation}
\label{diss_term_int}
    {D}(\varepsilon,{\bm r})=\frac{\pi }{2\varepsilon }\int\limits_{-\infty }^{\infty }{\int\limits_{0}^{\infty }{\bm \psi}_{1}({\bm x})\cdot {\bm S}({\bm r},\varepsilon {\bm x})\,{{x}_{r}}d{{x}_{r}}}d{{x}_{z}}\,.
\end{equation}
With a given velocity increment $\delta {\bm u}$, see eqns. \eqref{Velocity_dimless_BV}, and a test function from eq. \eqref{1_order_Gaussian_wavelet}, we can split ${\bm S}$ into a rotational, ${\bm S}_{\rm rot}=(\delta u_\varphi)^2\delta{\bm u}$, and an irrotational part, ${\bm S}_{\rm irr}=\left(\delta u_r^2+\delta u_z^2\right)\delta{\bm u}$. Then follow two terms for $D(\varepsilon,r)$ which are given by
\begin{align}
\label{diss_term_calc_ir}
   \frac{2{{D}_{\rm irr}(\varepsilon)}}{\pi {{\beta }_{v}}}&=\beta _{v}^{2}{{\varepsilon }^{2}}\int\limits_{-\infty }^{\infty }\exp \left( -x_{z}^{2}/2 \right)d{{x}_{z}}\int\limits_{0}^{\infty }x_{r}^{5}\exp \left( -x_{r}^{2}/2 \right)d{{x}_{r}} \\ 
 & +2\beta _{v}^{2}{{\varepsilon }^{2}}\int\limits_{-\infty }^{\infty }x_{z}^{2}\exp \left( -x_{z}^{2}/2 \right)d{{x}_{z}}\int\limits_{0}^{\infty }x_{r}^{3}\exp \left( -x_{r}^{2}/2 \right)d{{x}_{r}} \nonumber \\ 
 & -8\beta _{v}^{2}{{\varepsilon }^{2}}\int\limits_{-\infty }^{\infty }x_{z}^{4}\exp \left( -x_{z}^{2}/2 \right)d{{x}_{z}}\int\limits_{0}^{\infty }x_{r}\exp \left( -x_{r}^{2}/2 \right)d{{x}_{r}}\nonumber  \\ 
 \label{diss_term_calc_rot}
 \frac{2{{D}_{\rm rot}}(\varepsilon,r)}{\pi {{\beta }_{v}}}&=\int\limits_{-\infty }^{\infty }\exp \left( -x_{z}^{2}/2 \right)d{{x}_{z}}\int\limits_{0}^{\infty }x_{r}^{3}\exp \left( -x_{r}^{2}/2 \right)\delta u_{\varphi }^{2}d{{x}_{r}} \\ 
 & -2\int\limits_{-\infty }^{\infty }x_{z}^{2}\exp \left( -x_{z}^{2}/2 \right)d{{x}_{z}}\int\limits_{0}^{\infty }x_{r}\exp \left( -x_{r}^{2}/2 \right)\delta u_{\varphi }^{2}d{{x}_{r}}\nonumber 
\end{align}
The analytical solution of the integral for the irrotational part leads to a quadratic dependence on $\varepsilon$,
\begin{equation}
\label{diss_term_irrot}
    \frac{2D_{\rm irr}(\varepsilon)}{\pi\beta_v\sqrt{2\pi}}=-12\beta_v^2\varepsilon^2\,.
\end{equation}
For the rotational part, we obtain the following integral
\begin{equation}
\label{diss_term_rot}
    \frac{2D_{\rm rot}(\varepsilon,r)}{\pi\beta_v\sqrt{2\pi}}=\int\limits_{0}^{\infty }{{x_r}\left( x_{r}^{2}-2 \right)\exp \left( -\frac{x_{r}^{2}}{2} \right)\delta u_{\varphi }^{2}(r,\varepsilon x_r)\,d{{x}_{r}}}\,,
\end{equation}
which has to be evaluated numerically, it is shown in the Fig. \ref{Diss_term_contour}(a). In the panels (b)-(d) of the same figure, we sum both terms to obtain the total anomalous dissipation term $D(\varepsilon,r)$. The irrotational part of the dissipation term has a square dependency on the parameter $\beta_v$. For $\beta_v<0.1$ the rotational term has a strong influence on the full dissipation in the considered region, see panels (b) and (c) of the figure. For bigger $\beta_v$ (smaller Reynolds number) it can be seen that in the inertial range anomalous dissipation term varies to a good approximation with $\varepsilon^2$ as seen in panel (d). This results from the dominant irrotational term that is independent of $r$, see eq. \eqref{diss_term_irrot}. We observe in addition in the panel (d) of the figure that the variation of the isocontours of $D(\varepsilon,r)$ along the $r$-direction is practically absent for $r/r_B>2$. Since we have rescaled all distances in the contour plots by the Burgers radius $r_B$, the picture remains unchanged if we take the limit of $\nu\to 0$.

The cascade picture of turbulence suggests that energy is injected at a large scale. It is transferred to the smaller scales by the energy flux in the inertial range until it reaches the scale $\varepsilon/r_B=1$ where it starts to be transformed into heat. Scales of the size of the Burgers radius $r_B$ and smaller can be assigned to the viscous range in a turbulent flow (even though the present solution is a laminar and spatially smooth flow at all).  For filter scales in the inertial range, $\varepsilon>r_B$, the Burgers solution leads to a dissipation term that is always smaller at its center and larger at the core boundaries where the angular velocity reaches its maximum value. This effect disappears when the vortex circulation decreases $\Gamma \rightarrow 0$ which corresponds to $\beta_v \rightarrow\infty$. 

\begin{figure}[h]
\centering
\includegraphics[width=0.6\linewidth]{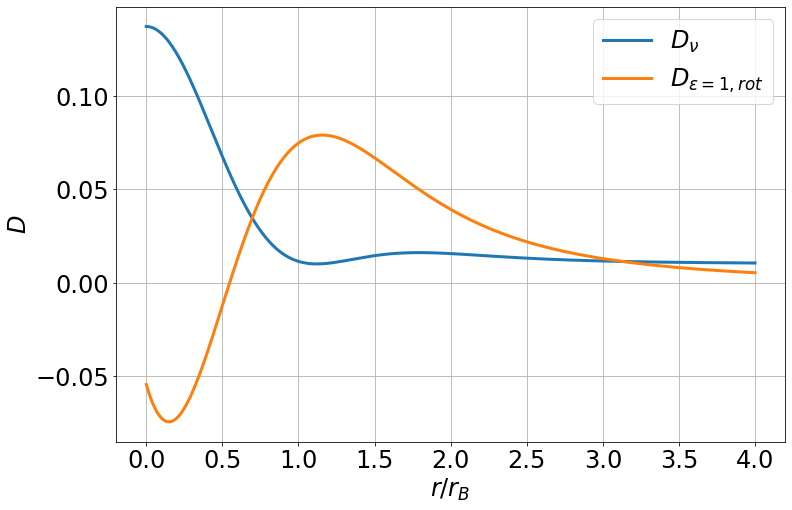}
\caption{Viscous dissipation term $D_{\nu}(r)$ in comparison with rotational part of the anomalous dissipation term $D_{\rm rot}(\varepsilon=1,r)$. While the viscous dissipation term has to be positive definite, this is not necessarily the case for the anomalous dissipation term.}
\label{Viscous_diss_figure}
\end{figure}

Finally, the anomalous dissipation term can be compared to the viscous dissipation term, which is given in dimensionless form for the kinematic model by $D_\nu=(\bm\nabla\bm u)^2/{\rm Re}_B$. Using the same form as in eqns. \eqref{diss_term_irrot} and \eqref{diss_term_rot} for a better comparison, we obtain 
\[\frac{2D_\nu }{\pi \beta _v\sqrt{2\pi }}=\frac{1}{\pi \sqrt{2\pi }}\left[ 5\beta _{v}^{2}+{{\left( 2\exp \left( -{{r}^{2}} \right)-\frac{1-\exp \left( -{{r}^{2}} \right)}{{{r}^{2}}} \right)}^{2}} \right]\,.\]
The corresponding distribution is shown in Fig. \ref{Viscous_diss_figure}. It is seen that the local anomalous dissipation can take negative and positive values while the viscous dissipation term is always positive definite. This is a major difference between both dissipation mechanisms. In the following, we increase the complexity of the vortex stretching model to further approach the situation in a turbulent Navier-Stokes flow. 

\section{Anomalous dissipation for randomly oriented Burgers vortex}
\subsection{Single randomly oriented Burgers vortex}
\begin{figure}[h]
\centering
\includegraphics[width=0.55\linewidth]{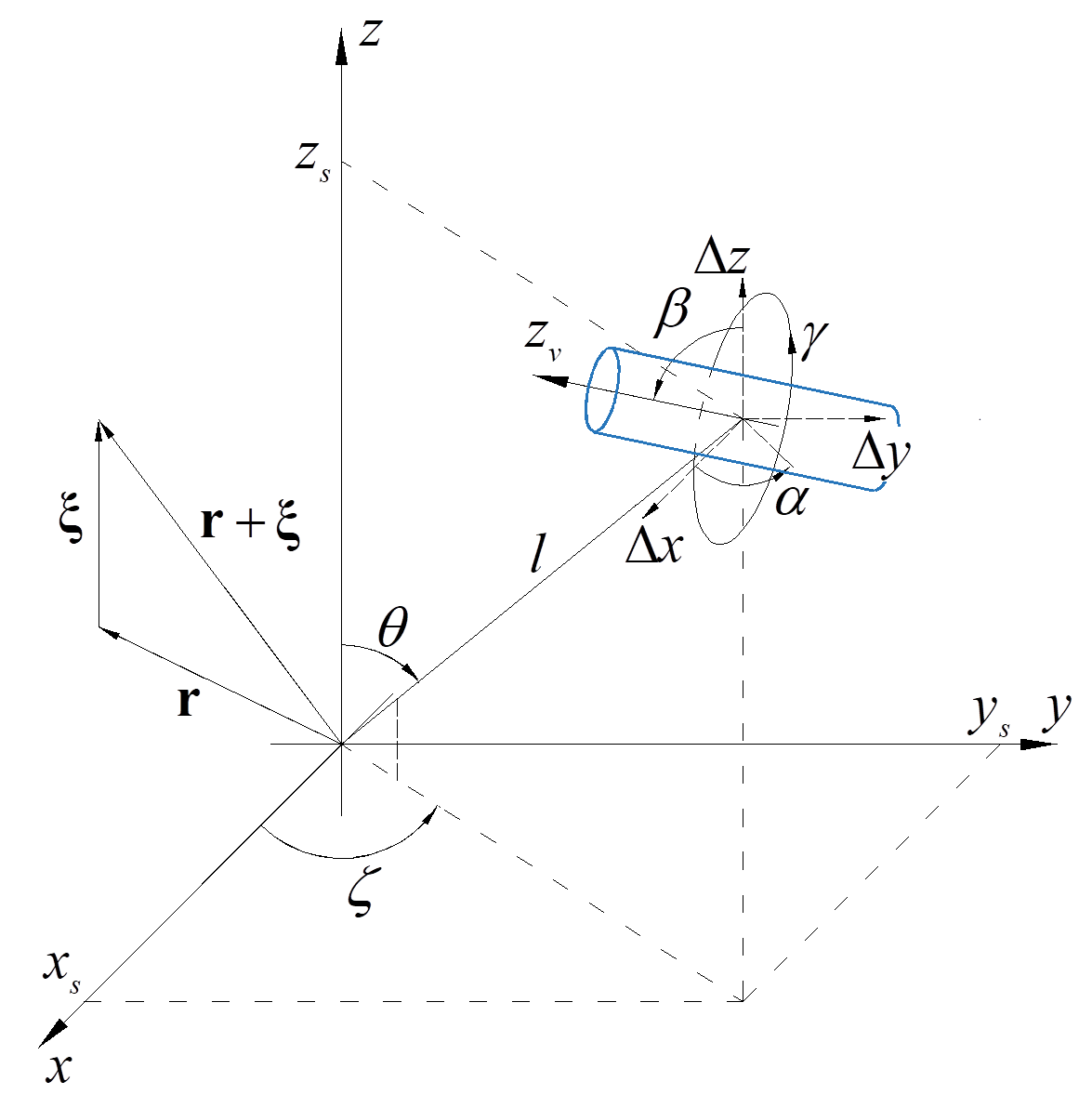}
\caption{Burgers vortex in the laboratory coordinate system. The parameters $( x_s,y_s,z_s)$ that can be written in spherical coordinates $(l,\zeta,\theta)$ determine the center point of the Burgers vortex. The angles $(\alpha,\beta,\gamma)$ determine the orientation of the vortex. Vector ${\bm r}=(x,y,z)$ represents the reference point and vector ${\bm \xi}=\left(0,0,\xi_z\right)$ represents the increment vector. The longitudinal velocity increment is defined as $\delta u_z=[{\bm u}({\bm r}+{\bm \xi})-{\bm u}({\bm r})]\cdot {\bm \xi}/|{\bm \xi}|=\delta_{||}u$.}
\label{vortex_in_labor_CC}
\end{figure}

\begin{figure}
    \centering
\includegraphics[width=1\textwidth]{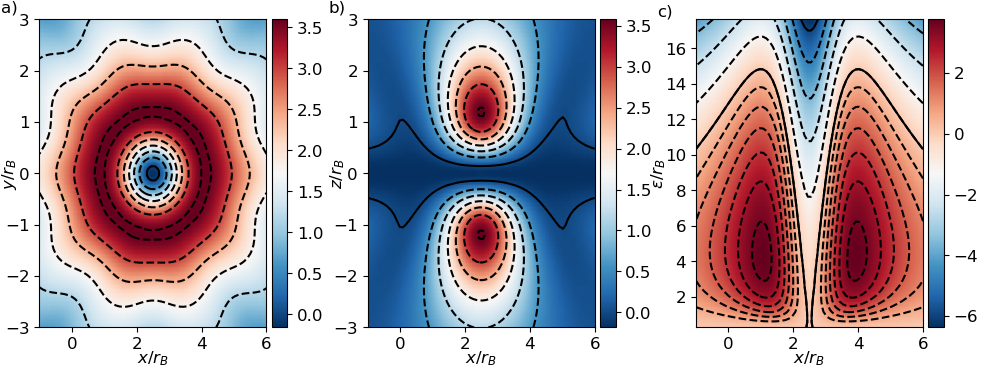}
\caption{Two-dimensional contour plots of the averaged anomalous dissipation term $\langle D_{\rm rot}(\varepsilon, r)+D_{\rm irr}(\varepsilon)\rangle_{\alpha,\beta}$ in units of ${u_0^3/r_B}$, amplified by a factor of $10^{-3}$, as given by eqns. \eqref{D_irrot_direction_average} and \eqref{D_rot_direction_average} for a single randomly oriented Burgers vortex in different planes. The vortex is located at $l/r_B=2.5;\ \theta=\pi /2;\ \zeta=0$. (a) $x$--$y$ cross section plane. (b) $x$--$z$ cross section plane for filter scale $\varepsilon/r_B=3$. (c) $x$--$\varepsilon$ plane. Since $l/r_B=2.5$ in the present example, the plots are symmetric with respect to $x/r_B=2.5$.}
\label{Diss_local_vortex}
\end{figure}

The results of the last section are now generalized. We therefore follow the ideas of Hatakeyama and Kambe for a kinematic model of turbulence in the form of randomly oriented Burgers vortices \cite{Hatakeyama1997,Kambe2000}. In the laboratory coordinate system, the center of the Burgers vortex is located at point $(x_s,y_s,z_s)$. These coordinates can be defined in terms of the spherical coordinates as follows, see Fig. \ref{vortex_in_labor_CC},
\begin{equation}
\label{shift}
    {{x}_{s}}  =l\cos \zeta \sin \theta,\quad {{y}_{s}}  =l\sin \zeta \sin \theta,\quad {{z}_{s}}  =l\cos \theta.
\end{equation}
A coordinate system with the axes $\Delta x,\Delta y,\Delta z$ is aligned  to the laboratory coordinate system and has its origin in the randomly oriented single Burgers vortex. Thus we have to rotate these coordinates by means of the angles $\alpha$, $\beta$, and $\gamma$, such that they get aligned with the local coordinate system which is centered along the inclined Burgers vortex axis at point $(x_v, y_v, z_v)$, see again Fig. \ref{vortex_in_labor_CC} for the illustration. The resulting rotation matrix $M$ which consists of three successive rotations is given by
\begin{equation}
\label{rotational_matrix}
    M={{R}_{z}}{{R}_{x}}{{R}_{z}}=\left( \begin{matrix}
   \cos \alpha \cos \gamma -\sin \alpha \cos \beta \sin \gamma  & -\cos \alpha \sin \gamma -\sin \alpha \cos \beta \cos \gamma  & \sin \alpha \sin \beta   \\
   \sin \alpha \cos \gamma +\cos \alpha \cos \beta \sin \gamma  & -\sin \alpha \sin \gamma +\cos \alpha \cos \beta \cos \gamma  & -\cos \alpha \sin \beta   \\
   \sin \beta \sin \gamma  & \sin \beta \cos \gamma  & \cos \beta  
\end{matrix} \right)\,.
\end{equation}
The local vortex coordinates $x_{i,v}=\left(x_v,y_v,z_v\right)=\left(r,\varphi,z_v\right)$ can be determined by the shifted coordinates $\Delta x_j$ and rotational matrix $M$ from \eqref{rotational_matrix}. This leads to
\begin{equation}
    {{x}_{i,v}}={{M}_{ij}}{\Delta x_j}\,.
\end{equation}
Connecting the cylindrical coordinates of the vortex with the shifted coordinate system, we obtain the following equations
\begin{equation}
\label{vortex_coord}
 {{x}_{v}}=r\cos \varphi ={{M}_{ix}}{\Delta x_i},\quad {{y}_{v}}=r\sin \varphi ={{M}_{iy}}{\Delta x_i} \quad \mbox{with}\quad  r^2 =x_{v}^{2}+y_{v}^{2}\,.
\end{equation}
With $\Delta x_i=x_i-x_{i,s}$ and eq. \eqref{shift}, we obtain additionally $z_v=M_{iz}(x_i-x_{i,s})$ and $r=[M_{ix}(x_i-x_{i,s})^2+M_{iy}(x_i-x_{i,s})^2]^{1/2}$. The increment vector ${\bm \xi}$ is then given by
\begin{equation}
\label{rel_ref_point}
\begin{aligned}
& {{x}'_{v}}-{{x}_{v}}={r}'\cos {\varphi }'-r\cos \varphi =\xi_z \sin \beta \sin \gamma,  \\ 
& {{y}'_{v}}-{{y}_{v}}={r}'\sin {\varphi }'-r\sin \varphi =\xi_z \sin \beta \cos \gamma , \\ 
& {{z}'_{v}}-{{z}_{v}}=\xi_z \cos \beta,  \\
&{{{r}'}^{2}}={{r}^{2}}+2r\xi_z \sin \left( \varphi +\gamma  \right)\sin \beta +\xi_z ^2{{\sin }^{2}}\beta .
\end{aligned}
\end{equation}
Here, parameters signed with an apostrophe refer to the point ${\bm r}+{\bm \xi}$. Velocity components can be connected to the laboratory coordinate system similarly. First, we switch from the cylindrical coordinate system to the Cartesian system
\begin{equation}
    {{u}_{x,v}}={{u}_{r}}\cos \varphi -{{u}_{\varphi }}\sin \varphi, \quad {{u}_{y,v}}={{u}_{r}}\sin \varphi +{{u}_{\varphi }}\cos \varphi, \quad {{u}_{z,v}}={{u}_{z,v}}\,.
\end{equation}
Then the rotation matrix \eqref{rotational_matrix} is used to obtain velocity components in the laboratory coordinate system
\begin{equation}
    {{u}_{j}}={{M}_{ji}}{{u}_{i,v}}\,.
\end{equation}
Compared to the model in Sec.~III, we consider velocity increments only along the longitudinal direction ${\bm \xi}$ which coincides with $z$-direction, i.e. ${\bm \xi}=(0,0,\xi_z)$. Then  $u_\xi=u_z$ in the laboratory coordinate system. Again, we separate the velocity into rotational and irrotational parts, which result to
\begin{equation}
\label{Velocity_labor_CS}
u_z=u_{z,{\rm rot}}+u_{z,{\rm irr}}
\end{equation}
with
\begin{align}
\label{Velocity_labor_CS_ir}
&u_{z,{\rm irr}}={{u}_{r}}\cos \varphi \sin \beta \sin \gamma +{{u}_{r}}\sin \varphi \sin \beta \cos \gamma +\cos \beta {{u}_{z,v}}, \\ 
\label{Velocity_labor_CS_rot}
&u_{z,{\rm rot}}=-{{u}_{\varphi }}\sin \varphi \sin \beta \sin \gamma +{{u}_{\varphi }}\cos \varphi \sin \beta \cos \gamma,
\end{align}
where the azimuthal angle is given by
\[\varphi =\arctan \left( \frac{{{y}_{v}}}{{{x}_{v}}} \right)=\arctan \left( \frac{{{M}_{iy}}{{x}_{i}}-{{M}_{iy}}{{x}_{i,s}}}{{{M}_{ix}}{{x}_{i}}-{{M}_{ix}}{{x}_{i,s}}} \right).\]
All velocity components in eqns. \eqref{Velocity_labor_CS_ir} and \eqref{Velocity_labor_CS_rot} are determined by \eqref{Velocity_dimless_BV}. The longitudinal velocity increment follows to
\begin{align}
\label{velocity_increment}
  & \delta {{u}_{z}}= [{\bm u}({\bm r}+{\bm \xi})-{\bm u}({\bm r})]\cdot \frac{{\bm \xi}}{|{\bm \xi}|}={{u}_{z}}\left( \mathbf{\Pi},x,y,z+\xi_z \right)-{{u}_{z}}\left( \mathbf{\Pi},x,y,z \right)\,,
\end{align}  
where $\mathbf{\Pi}=\left( \alpha ,\beta ,\gamma ,\theta ,\zeta ,l,\beta_v \right)$ determines the position and the direction of the vortex in the laboratory coordinate system, its circulation, strain rate, vortex radius and viscosity. By using eqns. \eqref{rel_ref_point}, \eqref{Velocity_labor_CS} and \eqref{velocity_increment}, the rotational and irrotational velocity increments follow to
\begin{align}
\label{velocity_increment_irrot}
  & \delta u_{z,{\rm irr}}={{\beta }_{v}}\xi \left( 3{{\cos }^{2}}\beta -1 \right) ,\\ 
  \label{velocity_increment_rot}
 & \delta u_{z,{\rm rot}}=\left( \frac{{{{{u}'}}_{\varphi }}}{{{r}'}}-\frac{{{u}_{\varphi }}}{r} \right)r\cos \left( \varphi +\gamma  \right)\sin \beta\,.
\end{align}
With the known velocity increment, the anomalous dissipation term can be calculated,
\begin{equation}
    D(\varepsilon,r)=\frac{1}{4{{\varepsilon }^{2}}}\int\limits_{-\infty }^{+\infty }{{\boldsymbol{\psi}}_{1}}\left( \frac{{\xi_z}}{\varepsilon } \right)\delta {u_z^3} {{d}\xi_z} \quad \mbox{with} \quad
    {{\psi }_{1}}( \xi_z)=-\xi_z \exp( -\xi_z^{2}/2)\,.
    \label{Full_integral_diss_term}
\end{equation}
Since the vortex is axisymmetric, we can set $\gamma=0$. With these simplification we can rewrite the velocity increment, the radial coordinate and the distance $r'$ to
\begin{equation}
\begin{aligned}
 \delta u_{z,{\rm irr}}&={{\beta }_{v}}\xi_z ( 3{{\cos }^{2}}\beta -1)\,,\\ 
\delta u_{z,{\rm rot}}&=\left( \frac{u^{\prime}_{\varphi}}{r^{\prime}}-\frac{u_{\varphi}}{r} \right)r\cos \varphi \sin \beta\,, \\ 
r'^2&={{r}^{2}}+2r\xi \sin  \varphi \sin \beta +{{\xi }^{2}}{{\sin }^{2}}\beta  ,\\ 
r^2&=(\cos \alpha \Delta x-\sin \alpha \cos \beta \Delta y+\sin \alpha \sin \beta \Delta z)^2+\\
&+(\sin \alpha \Delta x+\cos \alpha \cos \beta \Delta y-\cos \alpha \sin \beta \Delta z)^2.
\end{aligned}
\label{vel_incr_field_rot_irrot}
\end{equation}
In contrast to the case of the aligned Burgers vortex, the anomalous dissipation term is now determined for a randomly oriented Burgers vortex at a certain distance. The homogeneous isotropic statistics of the turbulence suggests thus an averaging over all directions of the vortex, i.e. over all possible angles $\alpha$ and $\beta$. We thus obtain
\begin{equation}
\begin{aligned}
   \langle D({\varepsilon },r)\rangle_{\alpha,\beta}&=\frac{1}{4{{\varepsilon }^{2}}}\frac{1}{4\pi }\int\limits_{-\infty }^{+\infty }{\int\limits_{0}^{\pi }{\int\limits_{0}^{2\pi }{{{\psi }_{1}}\left( \frac{\xi_z }{\varepsilon } \right)\delta u_{z,r}^{3}\sin\beta\, d\alpha }d\beta }d\xi_z } \\ 
 & +\frac{1}{4{{\varepsilon }^{2}}}\frac{1}{4\pi }\int\limits_{-\infty }^{+\infty }{\int\limits_{0}^{\pi }{\int\limits_{0}^{2\pi }{{{\psi }_{1}}\left( \frac{\xi_z }{\varepsilon } \right)3\delta u_{z,r}^{2}\delta {{u}_{z,ir}}\sin \beta\, d\alpha }d\beta }d\xi_z } \\ 
 & +\frac{1}{4{{\varepsilon }^{2}}}\frac{1}{4\pi }\int\limits_{-\infty }^{+\infty }{\int\limits_{0}^{\pi }{\int\limits_{0}^{2\pi }{{{\psi }_{1}}\left( \frac{\xi_z }{\varepsilon } \right)3\delta u_{z,r}\delta u_{z,ir}^{2}\sin \beta\, d\alpha }d\beta }d\xi_z } \\ 
 & +\frac{1}{4{{\varepsilon }^{2}}}\frac{1}{4\pi }\int\limits_{-\infty }^{+\infty }{\int\limits_{0}^{\pi }{\int\limits_{0}^{2\pi }{{{\psi }_{1}}\left( \frac{\xi_z }{\varepsilon } \right)\delta u_{z,ir}^{3}\sin \beta\, d\alpha }d\beta }d\xi_z. } 
\end{aligned}
\label{int1} 
\end{equation}
Integration over the $\alpha$ variable gives zero for the first and third terms in eq. \eqref{int1}. Therefore, the anomalous dissipation term can again be separated into a rotational part, the second integral term which is proportional to $\delta u_{z,{\rm rot}}^2\delta u_{z,{\rm irr}}$, and an irrotational part, the fourth integral term which is proportional to $\delta u_{z,{\rm irr}}^3$. The form of the integrals is similar to that of eqns. \eqref{diss_term_calc_ir} and \eqref{diss_term_calc_rot}. One gets now
\begin{align}
\label{D_irrot_direction_average}
    \frac{\langle {{D}_{\rm irr}(\varepsilon)}\rangle_{\alpha,\beta}}{{{\beta }_{v}}}&=-\frac{\beta _{v}^{2}}{8{{\varepsilon }^{3}}}\int\limits_{-\infty }^{+\infty }{\int\limits_{0}^{\pi }{{\xi_z^4}\exp \left( -\frac{{\xi_z ^2}}{2{{\varepsilon }^{2}}} \right){{( 3{{\cos }^{2}}\beta -1)}^{3}}\sin \beta\, d\beta }d\xi_z },\\
    \frac{\langle D_{\rm rot}(\varepsilon,\bm r)\rangle_{\alpha,\beta}}{{{\beta }_{v}}}&=\frac{3}{16\pi {\varepsilon ^3}}\int\limits_{-\infty }^{+\infty }{\int\limits_{0}^{\pi }{\int\limits_{0}^{2\pi }{{\xi_z ^2}\exp \left( -\frac{{\xi_z ^2}}{2{\varepsilon ^2}} \right)\left( \frac{u'_\varphi }{r'}-\frac{u_\varphi }{r} \right)^2r^2\cos ^2 \varphi \sin ^3\beta ( 1-3{{\cos }^{2}}\beta)\, d\alpha }d\beta }d\xi_z. }
    \label{D_rot_direction_average}
\end{align}
As a result, we receive for the irrotational part of the signal, a result that is similar to \eqref{diss_term_irrot}. It is given by 
\begin{equation}
    \frac{\langle D_{\rm irr}(\varepsilon)\rangle_{\alpha,\beta}}{{{\beta }_{v}}\sqrt{2\pi }}=-\frac{12}{35}\beta _{v}^{2}{{\varepsilon }^{2}}.
\end{equation}
The coefficient is different since we average over all possible vortex orientations and use the longitudinal velocity increment only in the present case. The distribution of full dissipation term is shown in Fig. \ref{Diss_local_vortex} for different cross section planes, see again Fig. \ref{vortex_in_labor_CC}. If we would set the distance from the coordinate origin to $l=0$ in panel (c) of the figure and thus move the vortex tube to the origin, we would get the same contours for shifted coordinates. Differences to the previous case in Sec. III near the center of the vortex arise due to the switch to longitudinal velocity increments. Similar to the aligned Burgers vortex case, the inertial flux vanishes as $\varepsilon ^2$ in the randomly oriented vortex case. Also at the boundaries of the Burgers vortex, the dissipation of energy reaches its maximum values; in the center it has the same smaller magnitude as further outside for all scales $\varepsilon$. In comparison to the previous case, energy dissipation does not have any inhomogeneities that are close to  the center of the Burgers vortex. But for the radial coordinate $r/r_B>0.5$ in the inertial range their behavior becomes similar.

\subsection{Ensemble of randomly oriented Burgers vortices}
\begin{figure}
    \centering
\includegraphics[width=1\textwidth]{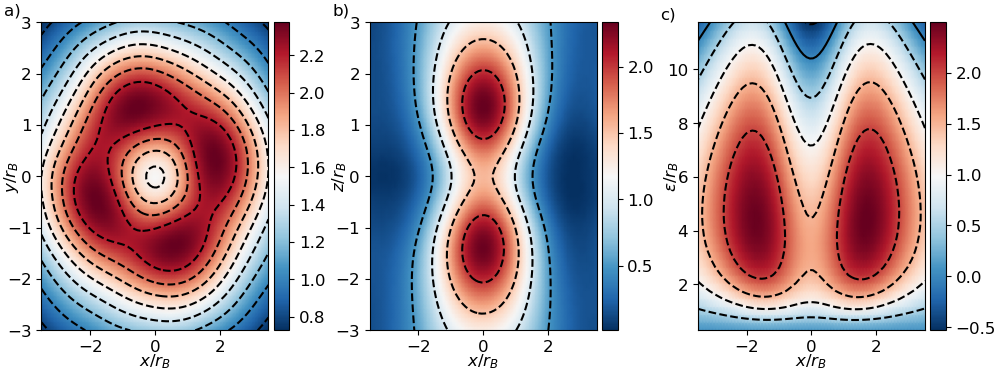}
\caption{Two-dimensional contour plots for the averaged anomalous dissipation term $\langle D_{\rm rot}(\varepsilon,r)+D_{\rm irr}(\varepsilon)\rangle_{\alpha,\beta,l,\theta,\zeta}$ in units of ${u_0^3/r_B}$, amplified by a factor of $10^{-3}$, given by eq. \eqref{D_rot_direction_position_average}. Differently to Fig. \ref{Diss_local_vortex}, the results are additionally averaged over $l$ and angles $\theta$ and $\zeta$ in correspondence with eq. \eqref{prob}. The contours are obtained for the most probable distance from the center of the laboratory coordinate system to the Burgers vortex of $l_0=1$, see eq. \eqref{prob}. (a) $x$--$y$ cross section plane.  (b) $x$--$z$ cross section plane for scale $\varepsilon/r_B=3$ and c) $x$--$\varepsilon$ plane.}
\label{Diss_vortex_averaged}
\end{figure}

If the center of the vortex tube is located randomly in a sphere of radius $l$ about the origin of the laboratory coordinate system, we have to average the dissipation term over angles $\theta$ and $\zeta$ additionally. Also the probability distribution of the distance $l$ is considered following ref. \cite{Kambe2000}. It is represented by the following probability density function
\begin{equation}
    P_{\rm vol}(V_l)=\frac{1}{\langle V\rangle}\exp{\left(-\frac{V_l}{\langle V\rangle}\right)}.
\end{equation}
Experimental evidence exists that a Poisson distribution of the spatial distribution of vortex filaments with respect to the origin is a good approximation \cite{Cadot1995}, as already discussed in  ref. \cite{Kambe2000}.  With the sphere volume $V_l=4\pi l^3/3$, the following distribution follows for $l$,
\begin{equation}
P_{\rm len}(l)=3bl^2\exp{(-bl^3)} \quad\mbox{with}\quad b=\frac{\Gamma(4/3)^3}{l_0^3}\,.
\label{prob}
\end{equation}
Thus, the dissipation term results to
\begin{equation}
\label{D_rot_direction_position_average}
    \langle D_{rot}(\varepsilon,x,y,z,\beta_v)\rangle_{\alpha,\beta,l,\theta,\zeta}=\frac{1}{4\pi}\int\limits_{0 }^{\infty }{\int\limits_{0}^{2\pi }{\int\limits_{0}^{\pi }\langle D_{rot}(\varepsilon,r)\rangle_{\alpha,\beta} P_l (l) \sin \theta d\theta }d\zeta }dl\,.
\end{equation}
The corresponding distribution is shown in Fig. \ref{Diss_vortex_averaged}. 
Adding a statistically random vortex  location does not affect the quadratic $\varepsilon$ dependence as it does not depend on its location. With the expected location of the vortex at $l_0=1$, the anomalous dissipation changes only very gradually within $r/r_B<l_0$, and exponentially decreases beyond it. For both statistical cases, the inhomogeneity of the energy dissipation is maximum in the inertial range at scales $\varepsilon/r_B=2-3$ and it is sign-positive in all $r-\varepsilon$ space while in the first case it changes it sign in the inertial range close to the vortex center.

\section{Local anomalous dissipation in isotropic turbulence}
\subsection{Direct numerical simulations}
As a last step in this analysis, we turn to an analysis of local and time-dependent anomalous dissipation events in a turbulent incompressible three-dimensional Navier-Stokes flow. This is done in direct numerical simulations of homogeneous isotropic box turbulence which solve the incompressible Navier-Stokes equations in a volume $V=L^3$ with periodic boundary conditions in all three spatial directions. We thus will analyse examples of high-vorticity events which result from an intense vortex stretching and investigate how close they are to the ideal scenarios that has been discussed in the last section. This implies a comparison with the single randomly oriented Burgers vortex.  

The Navier-Stokes equations of motion, see eqns. \eqref{NS_eq} and \eqref{C_eq}, are solved by a standard pseudospectral method. All fields are expanded in Fourier series, the switch between the physical and Fourier space is performed by fast Fourier transformations using the software package P3DFFT \cite{Pekurovsky2012}. The simulation domain is decomposed into pencils and the simulation code is parallelized with the Message Passing Interface. Time advancement is done by a second-order predictor-corrector scheme \cite{Schumacher2007,Schumacher2007a,Pushenko2024}.

The turbulence is in a statistically steady state which requires the addition of a volume forcing term ${\bm f}({\bm x},t)$ to the right hand side of eq. \eqref{NS_eq}. This forcing is defined such that at each time step a fixed amount of turbulent kinetic energy at a rate $\epsilon_{\rm in}$ is injected into the flow \cite{Schumacher2007,Schumacher2007a,Pushenko2024}. The forcing term is implemented in the Fourier space for the Fourier components $\hat{\bm u}({\bm k},t)$ at wavevector ${\bm k}$
\begin{equation}
    \label{eq:forcing}
    \mathfrak{F}\{{\bm f}({\bm x},t)\} = \epsilon_{\rm in} \frac{\hat{{\bm u}}({\bm k},t)}{\sum_{{\bm k}_f\in K} |\hat{\bm u}({\bm k}_f,t)|^2} \delta_{{\bm k},{\bm k}_f}\,.
\end{equation}
The subset of driven Fourier modes is given by $K = \{{\bm k}_f=(2\pi/L)(\pm 1,\pm 1, \pm 2)\}$ plus permutations of wavevector components. Here, $\epsilon_{\rm in}$ is the energy injection rate that prescribes the dissipation of turbulent kinetic energy, i.e.  $\epsilon_{\rm in} \approx \langle\epsilon\rangle$ for the statistically stationary regime. An additional parameter, which quantifies the strength of turbulence, the Taylor microscale Reynolds number which is given by
\begin{equation}
    R_\lambda = \sqrt{\frac{5}{3\nu \langle \epsilon \rangle}} U_{\rm rms}^2 \approx 10^2.
\end{equation}
In the present case, the simulation domain is covered by $N^3=1024^3$ grid points; the Kolmogorov dissipation length $\eta_K\approx 2a$ where the uniform grid spacing is $a=L/N$.

\subsection{Local anomalous dissipation for high-amplitude vorticity events}
\begin{figure}[t]
\centering
\includegraphics[width=0.95\linewidth]{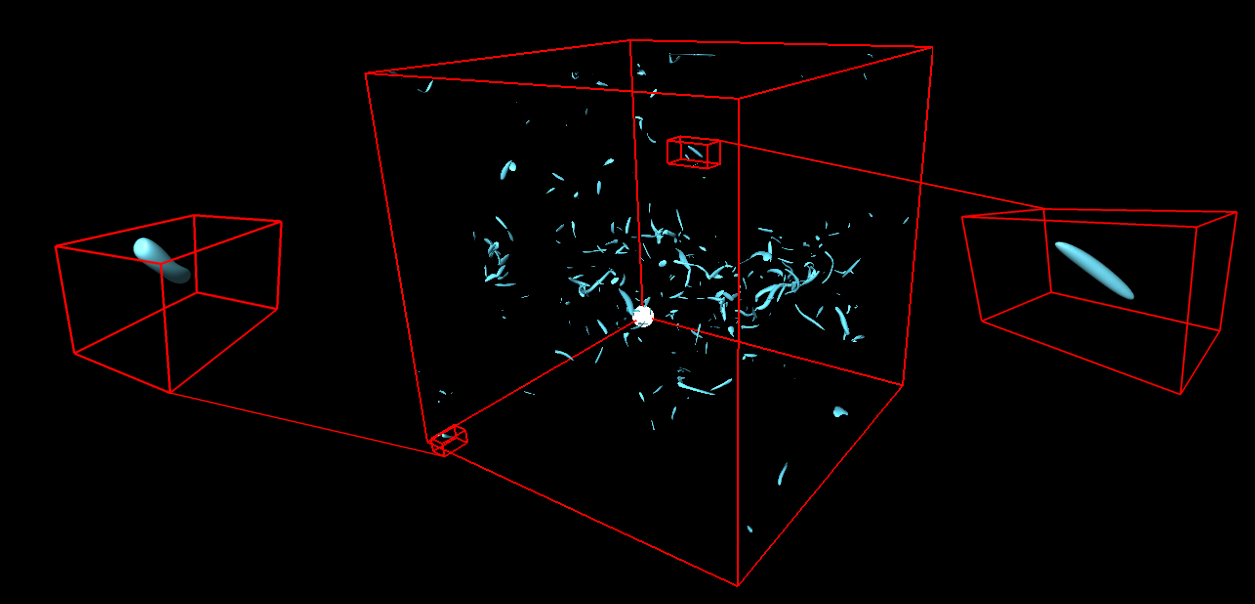}
\caption{Snapshot of the vorticity field taken from the DNS of homogeneous isotropic box turbulence. Vorticity isosurfaces $\mathbf{\omega}^2 \approx 100 \langle \omega^2\rangle_{V,t}$ are displayed where $\langle\cdot\rangle_{V,t}$ is a combined volume and time average. The simulation domain is resolved by a uniform mesh of $N^3=1024^3$ grid points. The two boxes to the left and right highlight two specific high-amplitude vorticity events. They are analysed in detail. The origin of the reference coordinate system is indicated by the white ball. Vortex tube examples 1 and 2 are shown on the right and left hande side magnifications, respectively.} 
\label{Vortex_tubes}
\end{figure}

In the following, we will pick two examples of high-vorticity events for the local analysis of precursors of anomalous dissipation. Therefore, we will match the local velocity field to that of the kinematic model of Sec. IV and compare the resulting field $D(\varepsilon,r)$. This implies that the local Burgers radii for both example cases correspond to $r_B \approx 3 \eta_K$. The isosurfaces of the vorticity field which highlight high-vorticity regions are presented in Fig. \ref{Vortex_tubes}. See also \cite{She1991} for a visualization of the corresponding field line bundles of the vorticity field. The two highlighted examples show tube-like isosurfaces with an inclined principal axis. 

\begin{figure}[h]
\centering
\includegraphics[width=0.95\linewidth]{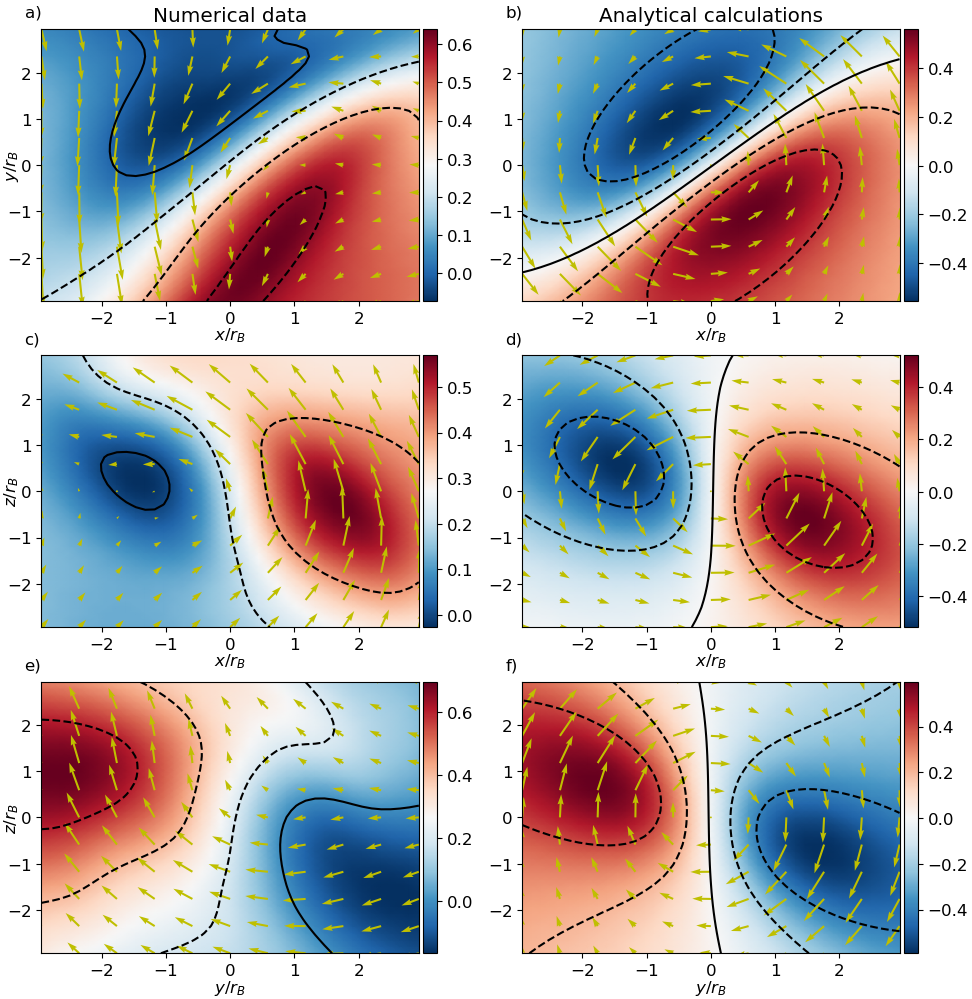}
\caption{Contour plot of the vertical velocity component $u_z/u_0$ in different planes for example 1. All plots are shifted to the center of the vortex. The difference between the two closest black contours is $0.2$. The solid line is the zero-contour-level in all panels. The yellow arrows stand for the corresponding velocity field projections.}
\label{Velocity_distrib}
\end{figure}
Recall that longitudinal increments are required only, which eases the analysis. To this end, we evaluate the anomalous dissipation over a finite interval of $z_0/\varepsilon \in [-8,8]$ with $z_0$ being the vertical coordinate of the center of the high-amplitude vortex tube event. Then the error of the numerical integration, which has to be applied here, is fallen below 0.1\%. In detail, we analyse the anomalous dissipation term locally for two individual events in the following with eq. \eqref{Full_integral_diss_term} and compare it to the analytical distribution of the anomalous dissipation term which we obtained in Sec. IV A. This is done (as said above) by taking the vertical velocity increment field \eqref{vel_incr_field_rot_irrot}. 

The coordinates of the center of example 1 of a stretched vortex tubes event (see the right zoom in Fig. \ref{Vortex_tubes}) is centered at $x_s=0.81 L,\ y_s=0.86 L,\ z_s=0.79 L$. According to Fig. \ref{vortex_in_labor_CC}, its spherical coordinates are $l=1.42 L,\ \zeta=0.814,\ \theta=0.983$ with the directional parameters $\alpha=-0.739,\ \beta= 1.045,\ \gamma=0$ (the result is independent of the parameter $\gamma$) as explained in Sec. IV. Note that the box with this event is close to the opposite corner of the origin along the diagonal. 

\begin{figure}[h]
\centering
\includegraphics[width=0.55\linewidth]{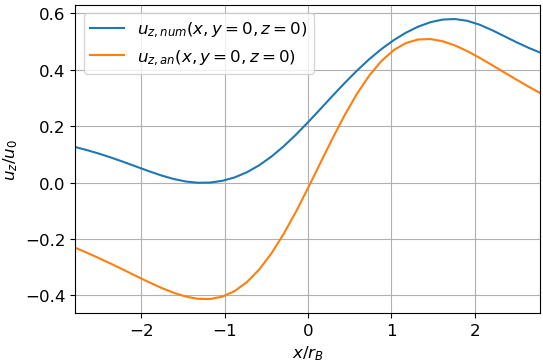}
\caption{Velocity profile $u_z/u_0$ along $x/r_B$-direction for example 1 with $u_0=\Gamma/(2\pi r_B)$, see \eqref{Velocity_labor_CS}. We compare numerical simulation data (num) and analytical model (an).}
\label{Vel_slice}
\end{figure}

In Fig. \ref{Velocity_distrib} the vertical velocity field component in different projection planes is compared for numerical data and analytical calculations. One can see that the analytical velocity distribution $u_z/u_0$ is always symmetric and its mean value is zero. In the 3D isotropic homogeneous flow, velocity maxima and minima are located at approximately the same positions, but its distribution is not necessarily symmetric;  it can be skewed to either positive or negative values which can be seen when comparing the color bars in the panels. One reason is that such a small-scale structure is typically swept on a larger-scale background vortex through the domain (even though this effect should be partly compensated for by the velocity increments which enter $D(\varepsilon,r)$).  Therefore, the slice cut in panels (c) and (d) of Fig. \ref{Velocity_distrib} is additionally shown in Fig. \ref{Vel_slice}. While the maximum amplitude in turbulence is of the same order as for the Burgers vortex model, the minimum amplitude is about zero. It can thus be concluded that the vertical velocity field component of the prototypical high-amplitude vorticity events in three-dimensional isotropic turbulence and of a randomly oriented Burgers vortex are qualitatively similar, but not identical. 

\begin{figure}[h]
\centering
\includegraphics[width=0.95\linewidth]{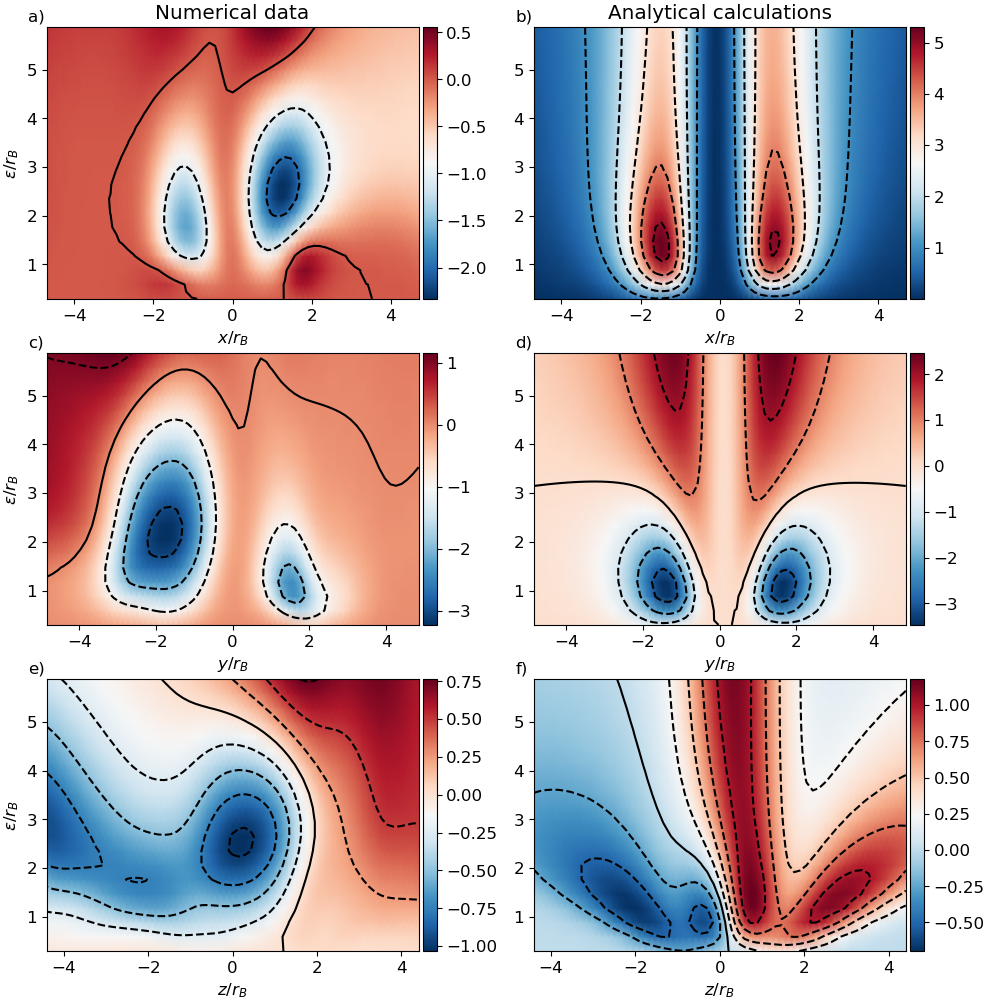}
\caption{Dissipation term $D/(u_0^3/r_B)$, amplified by a factor of $10^3$, in planes $x$--$\varepsilon$, $y$--$\varepsilon$, and $z$--$\varepsilon$ for the 3D isotropic homogeneous turbulence (example 1) and for the Burgers vortex. All contour planes are shown with respect to the center of the vortex. The solid lines in each panel stand for the zero-contour-level.}
\label{diss_1st_ev}
\end{figure}

Now the anomalous dissipation term can be calculated for example 1 using \eqref{Full_integral_diss_term}, for both, the numerical data and the analytical model based on the eqns. \eqref{velocity_increment_irrot} and \eqref{velocity_increment_rot} for the velocity increments. The result is shown in Fig. \ref{diss_1st_ev}. Even though the anomalous dissipation term in panels (a) and (b) has a different signs, it is qualitatively comparable. The different sign is caused by the difference in the velocity profiles. Panels (c) to (f) have similar magnitudes and display a qualitatively similar behaviour in the different cross-section planes. However for filter lengths $\varepsilon/r_B>3$, the deviations become bigger because other local stretching events appear in the vicinity of the selected subvolume. Another reason is that the Burgers vortex has an ideal velocity profile with a linear dependency of the radial and vertical velocity components: $u_r\sim r,\ u_z\sim z$ which cannot be found in a real, time-dependent homogeneous isotropic flow. 

\begin{figure}[h]
\centering
\includegraphics[width=0.95\linewidth]{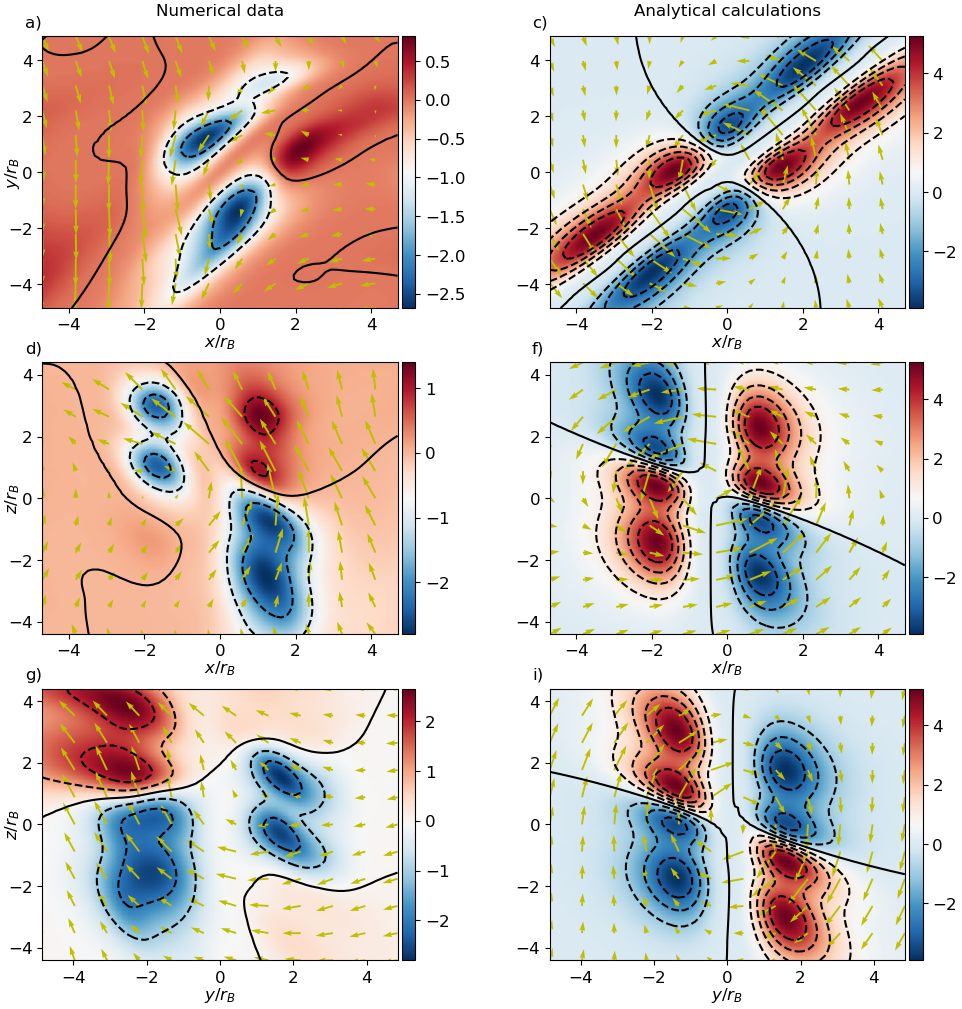}
\caption{Dissipation term $D/(u_0^3/r_B)$, amplified by a factor of $10^3$, in different planes for the 3D isotropic homogeneous turbulence (example 1) and for the Burgers vortex. Here $\varepsilon=1.2$. All contour planes are shown with respect to the center of the vortex. The solid lines in each panel stand for the zero-contour-level. The yellow arrows stand for the corresponding velocity field projections.}
\label{diss_1st_ev_with_vel_field}
\end{figure}

The dissipation term for a fixed $\varepsilon \approx 1.2$ plotted together with velocity fields (as vector arrays) can be found in the Fig. \ref{diss_1st_ev_with_vel_field} for different projection planes spanned by two of the three spatial coordinates $x$, $y$, $z$. It should be noted that the center of rotational motion of the vortex is located where the value of the dissipative term is approximately equal to zero $D\approx0$. Regions of higher anomalous dissipation magnitude can be identified in each panel; the structures are similar to those in the experimental results of ref. \cite{Dubrulle2019}. Furthermore, it is seen that the numerical data resemble those of the symmetric analytical model fairly well. 

\begin{figure}[h]
\centering
\includegraphics[width=0.95\linewidth]{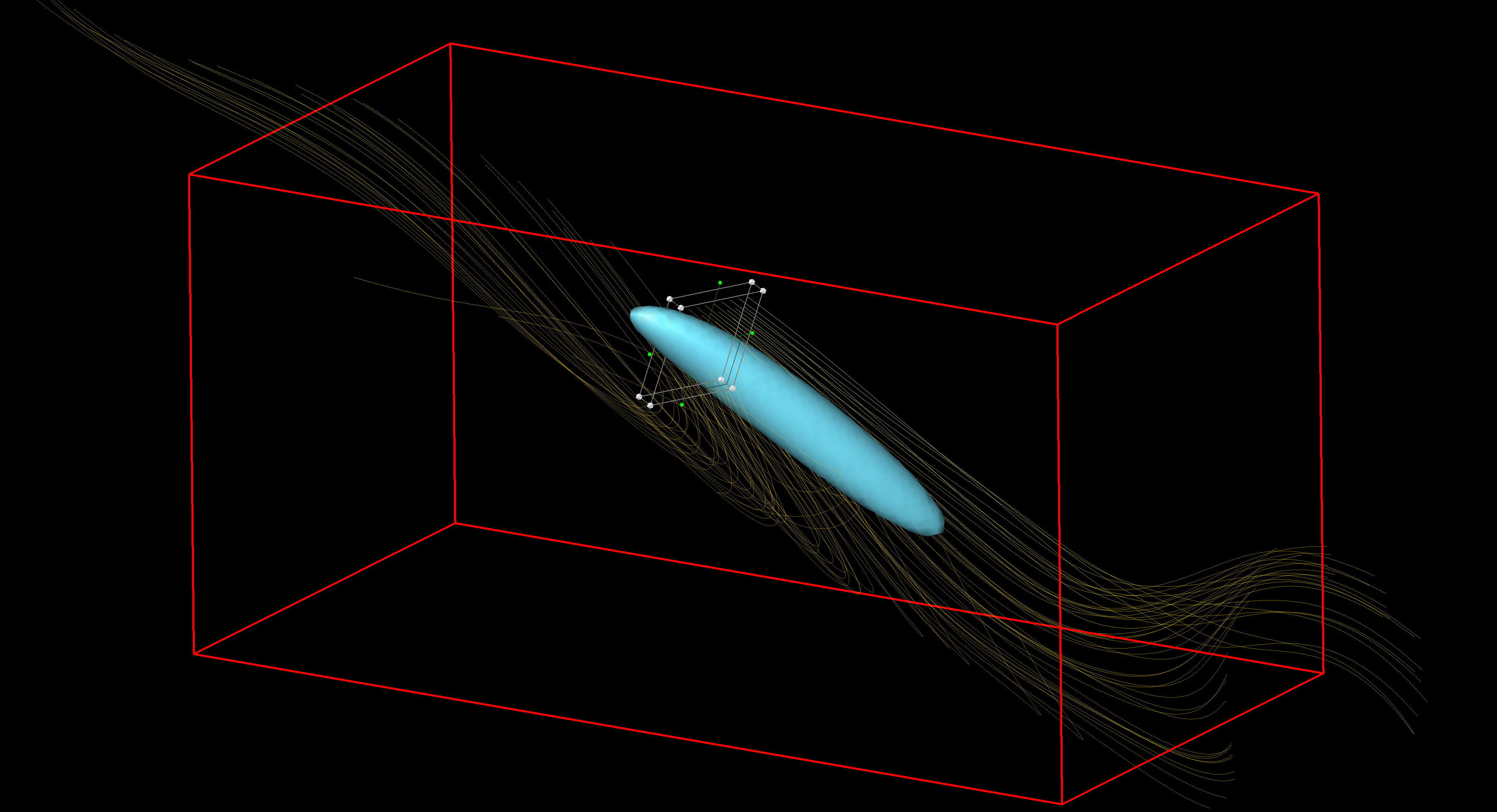}
\caption{Replot of the high-vorticity event example 1. The same high-vorticity magnitude isosurface as in Fig. \ref{Vortex_tubes} is shown together with the field lines of the vorticity vector field. They are seeded in the little white box that surrounds the vortex tube.}
\label{vortex_FL}
\end{figure}

A perfect match cannot be expected for the following additional reason. A Burgers vortex has a single vorticity component along its axis only, ${\bm \omega}=(0,0,\omega_{\parallel})$. The local turbulence events will display additional contributions perpendicular to the local vortex axis.
We suspect that this is a further source of deviations of the data from the ideal kinematic model. We find for example 1 the following relative magnitudes of the three vorticity components when the vector is centered at and aligned locally with the axis along the tube: $\omega^2_{\parallel}/\omega^2\approx 1$ and $\omega^2_{\perp}/\omega^2\approx 10^{-4}$ (for both perpendicular components). Fig. \ref{vortex_FL} animates the field lines around the high-vorticity event. It shows a good alignment with the local vertical axis, but diverging field lines further outside. 

\begin{figure}[h]
\centering
\includegraphics[width=0.95\linewidth]{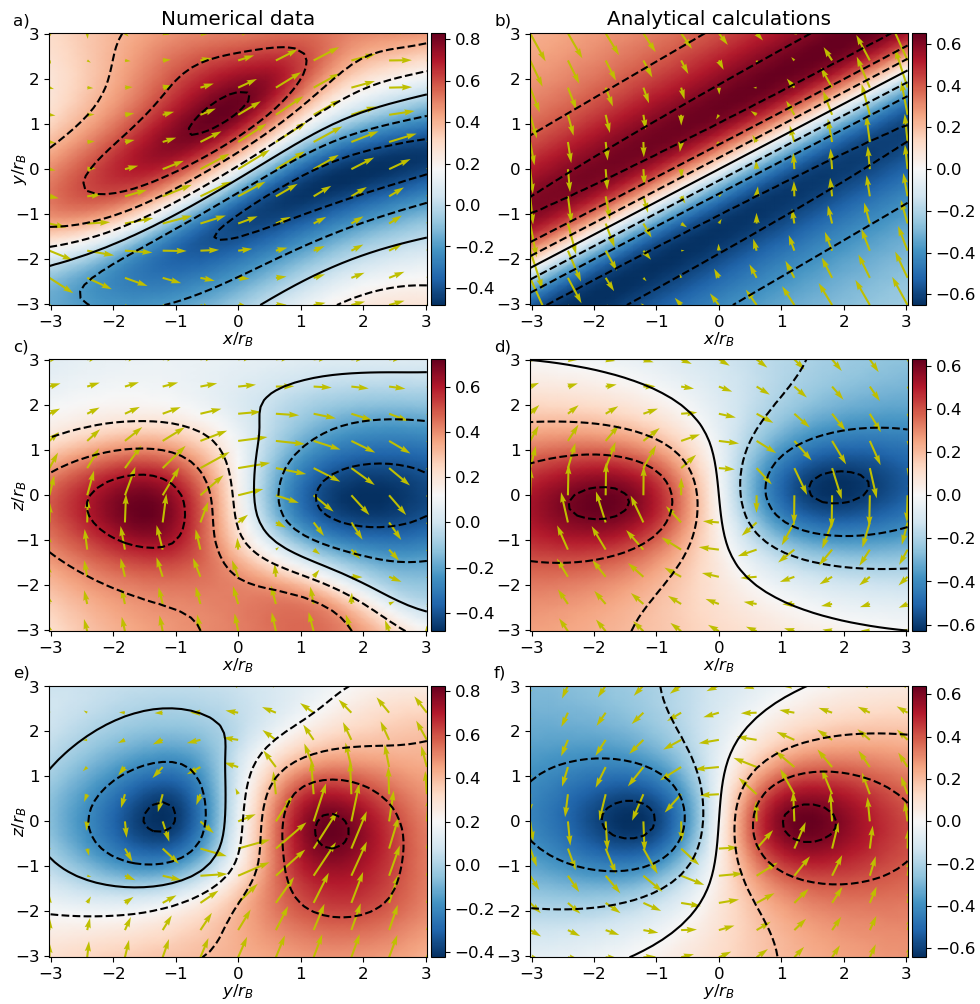}
\caption{Vertical velocity component distribution $u_z/u_0$ in different planes for example 2. All graphs are centralized to the center of the vortex. The difference between the two closest black contours is $0.2$. The solid line is the zero-contour-level in all panels. The yellow arrows stand for the corresponding velocity field projections.}
\label{Vel_2nd_ev}
\end{figure}

We have repeated the analysis for example 2 which was also highlighted in Fig. \ref{Vortex_tubes} (left box). The center of this second event is located at $x_s=0.96 L,\ y_s=0.13 L,\ z_s=0.03 L$. According to Fig. \ref{vortex_in_labor_CC}, its spherical coordinates are $l=0.96 L,\ \zeta=0.131,\ \theta=1.542$ with the directional parameters $\alpha=2.206,\ \beta= 1.511,\ \gamma=0$ (the result is independent of the parameter $\gamma$). The velocity field $u_z$ is shown in Fig. \ref{Vel_2nd_ev}. According to its different orientation to the previous event, the velocity field in different two-dimensional projection planes has another structure. 

\begin{figure}[h]
\centering
\includegraphics[width=0.95\linewidth]{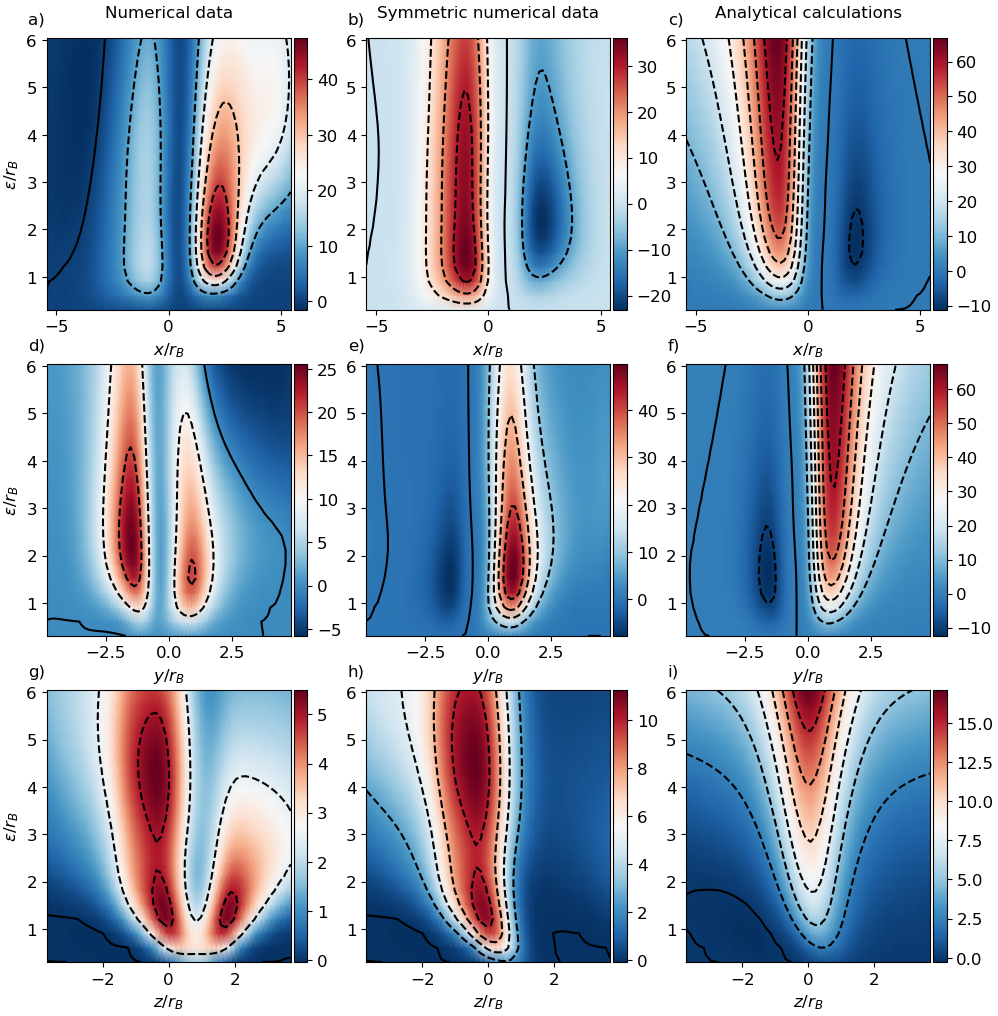}
\caption{Dissipation term $D/(u_0^3/r_B)$, amplified by a factor of $10^3$, in planes $x$--$\varepsilon$, $y$--$\varepsilon$, and $z$--$\varepsilon$ for the 3D isotropic homogeneous turbulence (example 2) and for the Burgers vortex. All contour planes are shown with respect to the center of the vortex. The solid lines in each panel stand for the zero-contour-level. The mid column of panels shows the results for the symmetrized DNS data.}
\label{diss_2nd_ev}
\end{figure}

The corresponding anomalous dissipation term is shown in Fig. \ref{diss_2nd_ev}. In contrast to the first event (example 1), the differences between numerical simulation results, see panels (a,d,g), and analytical calculations, see panels (c,f,i) of the figure, turn out to be more significantly. The reason of this difference is the even stronger asymmetry of the real vortex structure and the velocity field in its vicinity. In this case, we find the following relative magnitudes of the three vorticity components when the vector is centered at and aligned locally with the axis along the tube: $\omega^2_{\parallel}/\omega^2\approx 0.996$ and $\omega^2_{\perp}/\omega^2\approx 2\times 10^{-3}$ (for both perpendicular components). This event is thus slightly less aligned with the principal vortex tube axis.

\begin{figure}[h]
\centering
\includegraphics[width=1.0\linewidth]{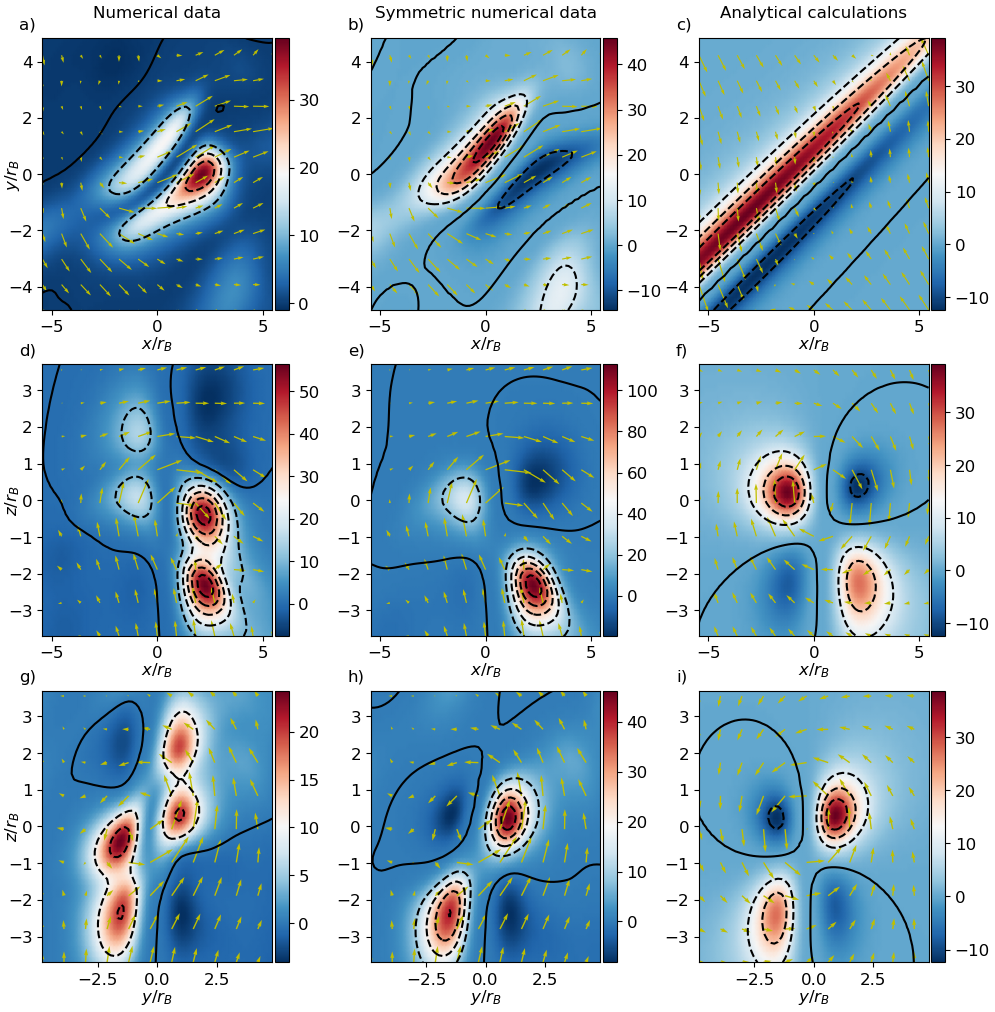}
\caption{Dissipation term $D/(u_0^3/r_B)$, amplified by a factor of $10^3$, in different planes for the 3D isotropic homogeneous turbulence (example 2) and for the Burgers vortex. All contour planes are shown with respect to the center of the vortex. The solid lines in each panel stand for the zero-contour-level. The velocity vector field is shown with yellow arrows. The mid column of panels shows the results for the symmetrized DNS data.}
\label{diss_2nd_ev_with_vel_field}
\end{figure}

For the vertical velocity increment it holds that $\delta u_z(r,|\xi|)\neq \delta u_z(r,-|\xi|)$. In contrast, for the ideal Burgers vortex, that we use for comparison, this equality is satisfied, i.e. $\delta u_z(r,|\xi|)= \delta u_z(r,-|\xi|)$. Therefore, we decided to `` symmetrize '' the real vortex configuration of the isotropic homogeneous turbulence locally by reflecting the velocity increment values from one side of the reference point $r$ to the opposite side. As a result of this operation and a corresponding calculation of the anomalous dissipation term for the symmetric real vortex, it appeared that the results qualitatively agree much better with those obtained for the Burgers vortex, panels (b,e,h) of Fig. \ref{diss_2nd_ev}. The dissipation term for a fixed $\varepsilon \approx 1.2$ with velocity fields can be found in the Fig. \ref{diss_2nd_ev_with_vel_field} for different two-dimensional projection planes, similar to Fig. \ref{diss_1st_ev_with_vel_field}. As for the first event, the center of rotational motion approximately corresponds to the minimum of the value of the dissipation term.


\section{Summary and Outlook}
The purpose of the present study was to identify precursors of anomalous dissipation in incompressible three-dimensional Navier-Stokes turbulence at finite Reynolds number. We have therefore enhanced the complexity of the underlying fluid flow model in incremental steps, starting with a simple steady Burgers vortex flow case which was generalized to a single randomly oriented Burgers vortex case and a corresponding kinematic model of a randomly distributed Burgers vortex ensemble afterwards. The latter two models go back to Kambe and Hatakeyama \cite{Hatakeyama1997,Kambe2000}. The results from these kinematic stretching models were eventually compared to direct numerical simulation data of three-dimensional, statistically steady, homogeneous isotropic turbulence in a cubic box with periodic boundary conditions in all three directions. In the latter case, we have extracted and analysed high-amplitude vorticity events in subvolumes of the full simulation domain; these are examples 1 and 2 which are magnified in Fig. \ref{Vortex_tubes}. 

The anomalous dissipation contribution $D(\varepsilon,r)$ was determined following the framework of Duchon and Robert \cite{Duchon2000}. Anomalous dissipation appears there as an additional term in the local kinetic energy balance and has to be taken to $\varepsilon\to 0$. We have investigated the magnitude dependence on the (finite) filter scale $\varepsilon$ and the distance $r$ from the origin, see again Fig. \ref{vortex_in_labor_CC}. We have shown that the results of the randomly oriented Burgers vortex model agree qualitatively well with those from the local analysis in a fully turbulent flow. Thus, we were able to bridge statistical behaviour -- as the anomalous dissipation -- to elementary structural building blocks of fluid turbulence -- local vortex stretching events. This was possible despite the fact that the present Reynolds numbers are moderate and the flow was not highly turbulent, to develop rough velocity fields in the inertial range. This is the reason of why consider our events as possible precursors to anomalous dissipation.

A possible extension of this study might be as follows: the Duchon-Robert framework can be extended to the fully compressible flow case where additional terms will appear due to the presence of dilatational velocity field fractions. This will affect the kinetic energy dissipation rate field, see e.g. \cite{John2021} for first attempts. Such studies are currently under way and will be reported elsewhere.

\acknowledgments
The work of GZ was supported by the Priority Programme DFG-SPP 2410 ``Hyperbolic Balance Laws in Fluid Mechanics: Complexity, Scales, Randomness (CoScaRa)`` funded by the Deutsche Forschungsgemeinschaft. It was also partly funded by the European Union (ERC, MesoComp, 101052786). Views and opinions expressed are however those of the author(s) only and do not necessarily reflect those of the European Union or the European Research Council. Neither the European Union nor the granting authority can be held responsible for them. The work of VP was supported by ITN CoPerMix. The project has received funding from the European Union’s Horizon 2020 research and innovation program under the Marie Skłodowska-Curie grant agreement N°956457. Supercomputing time has been provided at the University Computer Center (UniRZ) of the TU Ilmenau. The authors also gratefully acknowledge the Gauss Centre for Supercomputing e.V. (https://www.gauss-centre.eu) for funding this project by providing computing time on the GCS Supercomputer SuperMUC-NG at Leibniz Supercomputing Centre (https://www.lrz.de). 


\bibliographystyle{unsrt}

\begin{thebibliography}{10}

\bibitem{Kolmogorov1941}
A.~N. Kolmogorov.
\newblock {The Local Structure of Turbulence in Incompressible Viscous Fluid
  for Very Large Reynolds' Numbers}.
\newblock 30:301--305, 1941.

\bibitem{Frisch1995}
U.~Frisch.
\newblock {\em {Turbulence: The Legacy of A. N. Kolmogorov}}.
\newblock Cambridge University Press, Cambridge, UK, 1995.

\bibitem{Davidson2004}
P.~A. Davidson.
\newblock {\em Turbulence: An Introduction for Scientists and Engineers}.
\newblock Oxford University Press, Oxford, UK, 2004.

\bibitem{Dubrulle2019}
B.~Dubrulle.
\newblock {Beyond Kolmogorov cascades}.
\newblock {\em J. Fluid Mech.}, 867:P1, 2019.

\bibitem{Onsager1949}
L.~Onsager.
\newblock {Statistical hydrodynamics}.
\newblock {\em Nuovo Cimento Suppl.}, 6:{279--287}, 1949.

\bibitem{Eyink2024}
G.~Eyink.
\newblock {Onsager's ``Ideal Turbulence'' Theory}.
\newblock {\em J. Fluid Mech., submitted}, 2024.

\bibitem{Sreenivasan2024}
K.~R. Sreenivasan and J.~Schumacher.
\newblock {What is the turbulence problem, and when may we regard it as
  solved?}
\newblock {\em Annu. Rev. Condens. Matter Phys., submitted}, 2024.

\bibitem{Delellis2009}
C.~{De Lellis} and L.~{Sz\'{e}kelyhidi Jr.}
\newblock {The Euler equations as a differential inclusion}.
\newblock {\em Annal. Math.}, 170:{1417--1436}, 2009.

\bibitem{Isett2018}
P.~Isett.
\newblock {A proof of {O}nsager's conjecture}.
\newblock {\em Ann. Math.}, 188:871--963, 2018.

\bibitem{Sreenivasan1984}
K.~R. Sreenivasan.
\newblock On the scaling of the turbulence energy dissipation rate.
\newblock {\em Phys. Fluids}, 27:1048--1051, 1984.

\bibitem{Sreenivasan1998}
K.~R. Sreenivasan.
\newblock {An update on the energy dissipation rate in isotropic turbulence}.
\newblock {\em Phys. Fluids}, 10:528--529, 1998.

\bibitem{Kaneda2003}
Y.~Kaneda, T.~Ishihara, M.~Yokokawa, K.~Itakura, and A.~Uno.
\newblock {Energy dissipation rate and energy spectrum in high resolution
  direct numerical simulations of turbulence in a periodic box}.
\newblock {\em Phys. Fluids}, 15:{L21--L24}, 2003.

\bibitem{Pandey2022}
A.~Pandey, D.~Krasnov, K.~R. Sreenivasan, and J.~Schumacher.
\newblock {Convective mesoscale turbulence at very low Prandtl numbers}.
\newblock {\em J. Fluid Mech.}, 948:{A23}, 2022.

\bibitem{Burgers1948}
J.~M. Burgers.
\newblock A mathematical model illustrating the theory of turbulence.
\newblock {\em Adv. Appl. Mech.}, 1:171--199, 1948.

\bibitem{Townsend1951}
A.~A. Townsend.
\newblock On the fine-scale structure of turbulence.
\newblock {\em Proc. R. Soc. A}, 208:534--542, 1951.

\bibitem{Ashurst1987}
Wm.~T. Ashurst, A.~R. Kerstein, R.~M. Kerr, and C.~H. Gibson.
\newblock {Alignment of vorticity and scalar gradient with strain rate in
  simulated Navier–Stokes turbulence}.
\newblock {\em Phys. Fluids}, 30:2343--2353, 1987.

\bibitem{Ohkitani1994}
K.~Ohkitani.
\newblock {Kinematics of vorticity: Vorticity-strain conjugation in
  incompressible fluid flows}.
\newblock {\em Phys. Rev. E}, 50:5107--5110, 1994.

\bibitem{Hamlington2008a}
P.~E. Hamlington, J.~Schumacher, and W.~J.~A. Dahm.
\newblock {Local and nonlocal strain rate fields and vorticity alignment in
  turbulent flows}.
\newblock {\em Phys. Rev. E}, 77:{026303}, 2008.

\bibitem{Hamlington2008}
P.~E. Hamlington, J.~Schumacher, and W.~J.~A. Dahm.
\newblock {Direct assessment of vorticity alignment with local and nonlocal
  strain rates in turbulent flows}.
\newblock {\em Phys. Fluids}, 20:{111703}, 2008.

\bibitem{Duchon2000}
J.~Duchon and R.~Robert.
\newblock {Inertial energy dissipation for weak solutions of incompressible
  Euler and Navier-Stokes equations}.
\newblock {\em Nonlinearity}, 13:249--255, 2000.

\bibitem{Mallat1999}
S.~Mallat.
\newblock {\em A wavelet tour of signal processing}.
\newblock Academic Press, New York, 1999.

\bibitem{Hatakeyama1997}
N.~Hatakeyama and T.~Kambe.
\newblock Statistical laws of random strained vortices in turbulence.
\newblock {\em Phys. Rev. Lett.}, 79:1257--1260, 1997.

\bibitem{Kambe2000}
T.~Kambe and N.~Hatakeyama.
\newblock Statistical laws and vortex structures in fully developed turbulence.
\newblock {\em Fluid Dyn. Res.}, 27:247--267, 2000.

\bibitem{Saw2016}
E.-W. Saw, D.~Kuzzay, D.~Faranda, A.~Guittonneau, F.~Daviaud,
  C.~Wiertel-Gasquet, V.~Padilla, and B.~Dubrulle.
\newblock {Experimental characterization of extreme events of inertial
  dissipation in a turbulent swirling flow}.
\newblock {\em Nat. Commun.}, 7:{12466}, 2016.

\bibitem{Pope2000}
S.~B. Pope.
\newblock {\em {Turbulent Flows}}.
\newblock Cambridge University Press, Cambridge, UK, 2000.

\bibitem{Cadot1995}
O.~Cadot, S.~Douady, and Y.~Couder.
\newblock Characterization of the low‐pressure filaments in a
  three‐dimensional turbulent shear flow.
\newblock {\em Phys. Fluids}, 7:630--646, 1995.

\bibitem{Pekurovsky2012}
D.~Pekurovsky.
\newblock {P3DFFT: A} framework for parallel computations of fourier transforms
  in three dimensions.
\newblock {\em SIAM J. Sci. Comput.}, 34:C192--C209, 2012.

\bibitem{Schumacher2007}
J.~Schumacher, K.~R. Sreenivasan, and V.~Yakhot.
\newblock {Asymptotic exponents from low-Reynolds-number flows}.
\newblock {\em New J. Phys.}, 9:{89}, 2007.

\bibitem{Schumacher2007a}
J.~Schumacher.
\newblock {Sub-Kolmogorov-scale fluctuations in fluid turbulence}.
\newblock {\em Europhys. Lett.}, 80:{54001}, 2007.

\bibitem{Pushenko2024}
V.~Pushenko and J.~Schumacher.
\newblock Connecting finite-time lyapunov exponents with supersaturation and
  droplet dynamics in a turbulent bulk flow.
\newblock {\em Phys. Rev. E}, 109:045101, 2024.

\bibitem{She1991}
Z.-S. She, E.~Jackson, and S.~A. Orszag.
\newblock Intermittent vortex structures in homogeneous isotropic turbulence.
\newblock {\em Nature}, 344:226--228, 1991.

\bibitem{John2021}
J.~Panickacheril John, D.~A. Donzis, and K.~R. Sreenivasan.
\newblock Does dissipative anomaly hold for compressible turbulence?
\newblock {\em J. Fluid Mech.}, 920:A20, 2021.

\end{thebibliography}

\end{document}